\pdfoutput=1
\documentclass[usenatbib]{mn2e}
\usepackage{apjfonts}
\usepackage{amssymb}
\usepackage{amsmath}
\usepackage{ctable}
\usepackage{url}
\usepackage{fixltx2e} 

\newcommand{\be}{\begin{equation}}
\newcommand{\ee}{\end{equation}}

\newcommand{\etal}{et al.}

\newcommand{\msun}{M_{\sun}}

\newcommand{\paperone}{Paper {\small I}}
\newcommand{\papertwo}{Paper {\small II}}
\newcommand{\paperthree}{Paper {\small III}}
\newcommand{\paperfour}{Paper {\small IV}}

\newcommand{\movieurl}{\url{http://www.tapir.caltech.edu/~phopkins/Site/Movies_sbw_mgr.html}}

\newcommand{\scaleup}{}

\newcommand\plotonesize[2]
 {\centering \leavevmode \includegraphics[width={#2\columnwidth}]{#1}}
\newcommand{\plotsidesize}[2]
 {\centering \leavevmode \includegraphics[width={#2\textwidth}]{#1}}
\newcommand{\acknowledgments}{\begin{small}\section*{Acknowledgments}\end{small}}
\newcommand\altaffilmark[1]{$^{#1}$}
\newcommand\altaffiltext[1]{$^{#1}$}
\title[Starburst Winds \&\ Galaxy Mergers]{Resolving The Generation of Starburst Winds in Galaxy Mergers\vspace{-0.5cm}
}

\author[Hopkins \etal]{
\parbox[t]{\textwidth}{ 
Philip F.~Hopkins\thanks{E-mail:phopkins@caltech.edu}\altaffilmark{1,2},
Dusan Kere\v{s}\altaffilmark{3}, 
Norman Murray\altaffilmark{4,5}, 
Lars Hernquist\altaffilmark{6}, \\
Desika Narayanan\altaffilmark{7}, \&\ 
Christopher C.~Hayward\altaffilmark{8}
} 
\vspace*{6pt} \\
\altaffiltext{1}{TAPIR, Mailcode 350-17, California Institute of Technology, Pasadena, CA 91125, USA} \\
\altaffiltext{2}{Department of Astronomy and Theoretical Astrophysics Center, University of California Berkeley, Berkeley, CA 94720} \\
\altaffiltext{3}{Department of Physics, Center for Astrophysics and Space Science, University of California at San Diego, 9500 Gilman Drive, La Jolla, CA 92093} \\ 
\altaffiltext{4}{Canadian Institute for Theoretical Astrophysics, 
60 St.\ George Street, University of Toronto, ON M5S 3H8, Canada} \\
\altaffiltext{5}{Canada Research Chair in Astrophysics} \\
\altaffiltext{6}{Harvard-Smithsonian Center for Astrophysics, 60 
Garden Street, Cambridge, MA 02138} \\
\altaffiltext{7}{Steward Observatory, University of Arizona, 933 
N Cherry Ave, Tucson, Az, 85721} \\ 
\altaffiltext{8}{Heidelberger Institut f\"ur Theoretische Studien, Schlo\ss-Wolfsbrunnenweg 35, 69118 Heidelberg, Germany} 
\vspace{-0.5cm}
}

\date{Submitted to MNRAS, December, 2012\vspace{-0.5cm}}
\begin{document}
\maketitle
\label{firstpage}

\begin{abstract}

We study galaxy super-winds driven in major mergers, using pc-scale resolution simulations with detailed models for stellar feedback that can self-consistently follow the generation of winds. The models include molecular cooling, star formation at high densities in GMCs, and gas recycling and feedback from SNe (I \&\ II), stellar winds, and radiation pressure. We study mergers of systems from SMC-like dwarfs and Milky Way analogues to $z\sim2$ starburst disks. Multi-phase super-winds are generated in all passages, with outflow rates up to $\sim1000\,M_{\sun}\,{\rm yr^{-1}}$. However, the wind mass-loading efficiency (outflow rate divided by star formation rate) is similar to that in the isolated galaxy counterparts of each merger: it depends more on global galaxy properties (mass, size, and escape velocity) than on the dynamical state or orbital parameters of the merger. Winds tend to be bi- or uni-polar, but multiple `events' build up complex morphologies with overlapping, differently-oriented bubbles and shells at a range of radii. The winds have complex velocity and phase structure, with material at a range of speeds up to $\sim1000\,{\rm km\,s^{-1}}$ (forming a Hubble-like flow), and a mix of molecular, ionized, and hot gas that depends on galaxy properties. We examine how these different phases are connected to different feedback mechanisms. These simulations resolve a problem in some ``sub-grid'' models, where simple wind prescriptions can dramatically suppress merger-induced starbursts, often making it impossible to form ULIRGs. Despite large mass-loading factors ($\gtrsim10-20$) in the winds simulated here, the peak star formation rates are comparable to those in ``no wind'' simulations. Wind acceleration does not act equally, so cold dense gas can still lose angular momentum and form stars, while these stars blow out gas that would not have participated in the starburst in the first place. Considerable wind material is not unbound, and falls back on the disk at later times post-merger, leading to higher post-starburst SFRs in the presence of stellar feedback. We consider different simulation numerical methods and their effects on the wind phase structure; while most results are converged, we find that the existence of small clumps in the outflow at large distances from the galaxy is quite sensitive to the methodology.

\end{abstract}

\begin{keywords}
galaxies: formation --- galaxies: evolution --- galaxies: active --- 
star formation: general --- cosmology: theory\vspace{-0.5cm}
\end{keywords}

\vspace{-0.5cm}
\section{Introduction}
\label{sec:intro}

It is well-established that feedback from stars is a key component of galaxy formation models. Absent strong stellar feedback, gas in cosmological models quickly cools and turns into stars, predicting galaxies with much larger stellar masses than observed \citep[e.g.][and references therein]{katz:treesph,somerville99:sam,cole:durham.sam.initial,
springel:lcdm.sfh,keres:fb.constraints.from.cosmo.sims}. ``Slowing down'' star formation does not eliminate this problem; the real issue is that the amount of baryons in real galactic disks is much lower than the universal baryon fraction, which is the predicted amount of gas and stars found in cosmological simulations of low-mass galaxies without strong feedback
(\citealt{white:1991.galform}; for a recent review see \citealt{keres:fb.constraints.from.cosmo.sims}). Observational constraints from IGM enrichment further make clear that many of those baryons must have at one point entered galaxy halos and disks, and been enriched, then ejected 
\citep{aguirre:2001.igm.metal.evol.sims,pettini:2003.igm.metal.evol,songaila:2005.igm.metal.evol,martin:2010.metal.enriched.regions}. Galactic super-winds are therefore implied, with large mass-loading factors of several times the SFR that are required in cosmological simulations to reproduce these observations \citep[e.g.][]{oppenheimer:outflow.enrichment}. Such mass-loading factors are also observationally inferred in many local galaxies and massive star-forming regions at $z\sim2-3$ \citep{martin99:outflow.vs.m,
martin06:outflow.extend.origin,heckman:superwind.abs.kinematics,newman:z2.clump.winds,sato:2009.ulirg.outflows,chen:2010.local.outflow.properties,
steidel:2010.outflow.kinematics,coil:2011.postsb.winds}.

Until recently, however, numerical simulations have generally been unable to produce, from an a priori model, winds with large mass-loading factors (as well as a plausible scaling of wind mass-loading with galaxy mass or other properties); this is especially true of models which include only thermal or ``kinetic'' feedback via supernovae, which is very inefficient in the dense regions where star formation occurs \citep[see e.g.][and references therein]{guo:2010.hod.constraints,powell:2010.sne.fb.weak.winds,brook:2010.low.ang.mom.outflows,nagamine:2010.dwarf.gal.cosmo.review,bournaud10}. More recent simulations have, with higher resolution and/or stronger feedback prescriptions, seen strong winds, but generally find it is critical to include (usually simplified) prescriptions for cooling suppression and/or ``pre-supernovae'' feedback \citep[see][]{governato:2010.dwarf.gal.form,maccio:2012.cuspcore.outflows,teyssier:2013.cuspcore.outflow}. This should not be surprising: feedback processes other than supernovae are critical for suppressing star formation in dense gas; these include protostellar jets, HII photoionization, stellar winds, and radiation pressure from young stars. Including  these mechanisms self-consistently maintains a reasonable fraction of the ISM at densities where the thermal heating from supernovae has a larger effect; moreover there are many regimes where these mechanisms can directly drive winds, independent of and with greater mass loading than supernovae.  

This conclusion implies that (not surprisingly) an accurate treatment of galactic winds requires a more realistic treatment of the stellar feedback processes that maintain the multi-phase structure of the ISM of galaxies. Motivated by these problems, in \citet{hopkins:rad.pressure.sf.fb} (\paperone) and \citet{hopkins:fb.ism.prop} (\papertwo), we developed a new set of numerical models to follow feedback on small scales in GMCs and star-forming regions, in simulations with pc-scale resolution.\footnote{\label{foot:url}Movies of these
  simulations are available at \movieurl} 
These simulations include the momentum imparted locally (on sub-GMC 
scales) from stellar radiation pressure, radiation pressure on larger scales via the light that escapes star-forming regions, HII photoionization heating, as well as the heating, momentum deposition, and mass loss by SNe (Type-I and Type-II)  and stellar winds (O star and AGB).  The feedback is tied to the young stars, with the energetics and time-dependence taken directly from stellar evolution models. Our models also include cooling to temperatures $<100\,$K, and a treatment of the molecular/atomic transition in gas and its effect on star formation \citep[following][]{krumholz:2011.molecular.prescription}. We showed that these feedback mechanisms produce a quasi-steady ISM in which giant molecular clouds form and disperse rapidly, after turning just a few percent of their mass into stars.   This leads to an ISM with phase structure, turbulent velocity dispersions, scale heights, and GMC properties (mass functions, sizes, scaling laws) in reasonable agreement with observations.

In \citet{hopkins:stellar.fb.winds} (\paperthree), we show that these same models of stellar feedback {\em predict} the elusive winds invoked in almost all galaxy formation models; the {\em combination} of multiple feedback mechanisms is critical to give rise to massive, multi-phase winds having a broad distribution of velocities, with material both stirred in local fountains and unbound from the disk. 

However, in \paperthree\ we examine only idealized isolated disk galaxies. Although this is probably representative of much of a galaxy's lifetime, a great deal of observational study has focused on winds in ``starburst'' galaxies, often in interacting or merging systems. Indeed, a wide range of phenomena indicate that gas-rich mergers are important to galaxy formation and star formation. These systems dominate the most intense starburst populations: ULIRGs at low redshift \citep{joseph85,sanders96:ulirgs.mergers}, and Hyper-LIRGs and bright sub-millimeter galaxies at high redshifts \citep{papovich:highz.sb.gal.timescales,
younger:smg.sizes,tacconi:smg.maximal.sb.sizes,
schinnerer:submm.merger.w.compact.mol.gas,
chapman:submm.halo.clustering,tacconi:smg.mgr.lifetime.to.quiescent}. They are powered by compact concentrations of gas at enormously high densities \citep{scoville86,sargent87}, which provides material to fuel BH growth and boost the concentration and central phase-space density of merging disks to match those of ellipticals \citep{hernquist:phasespace,robertson:fp,hopkins:cusps.mergers}. Various studies have shown that the mass involved in these starburst events 
is critical for explaining the relations between spirals, mergers, and ellipticals, 
and has a dramatic impact on the properties of merger 
remnants \citep[e.g.,][]{LakeDressler86,Doyon94,ShierFischer98,James99,
Genzel01,tacconi:ulirgs.sb.profiles,dasyra:mass.ratio.conditions,dasyra:pg.qso.dynamics,rj:profiles,rothberg.joseph:kinematics,hopkins:cusps.ell,hopkins:cores}. 

With central densities as large as $\sim1000$ times those in Milky Way giant molecular clouds (GMCs), these systems also provide a laboratory for studying star formation, the ISM, and the generation of galactic winds under the most extreme conditions. In \citet{hopkins:stellar.fb.mergers} (\paperfour), we therefore extend the models from \paperone-\paperthree\ to include major galaxy mergers. We showed there that the same feedback mechanisms can explain the self-regulation of starbursts and extension of the Kennicutt-Schmidt relation to the highest gas surface densities observed. We also show how this controls the star formation rates and their spatial distributions, the formation of clusters, and the formation and destruction of giant molecular clouds in the ISM. In this paper, we further investigate the phase structure and generation of galactic superwinds in these models, and how they relate to merger dynamics and star formation histories.

\begin{figure}
    \centering
    \plotonesize{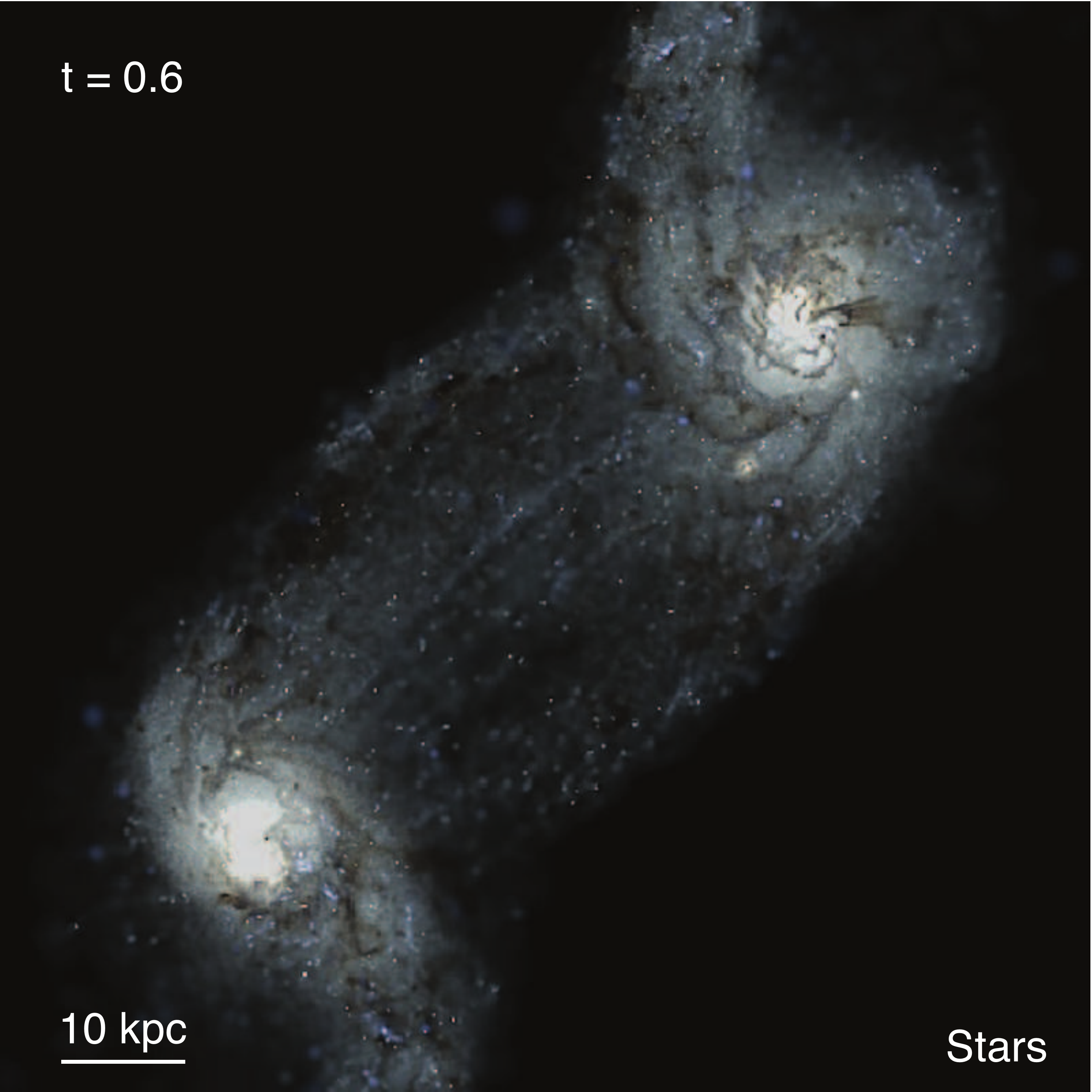}{1}
    \caption{Morphology of a standard simulation (all feedback mechanisms included) of a merger of the HiZ disk model (a massive, $z\sim2-4$ starburst disk merger). The time is near apocenter after first passage. 
    The image is a mock $ugr$ (SDSS-band) composite, 
    with the spectrum of all stars calculated from their known age and metallicity, 
    and dust extinction/reddening accounted for from the line-of-sight dust mass. 
    The brightness follows a logarithmic scale with a stretch of $\approx2\,$dex. Young 
    star clusters are visible throughout the system as bright white pixels. 
    The nuclei contain most of the star formation, but considerable fine structure in the dust and gas gives rise to complicated filaments, dust lanes, and patchy obscuration of star-forming regions. 
    \label{fig:morph.demo.stars}}
\end{figure}

\begin{figure*}
    \centering
    \plotsidesize{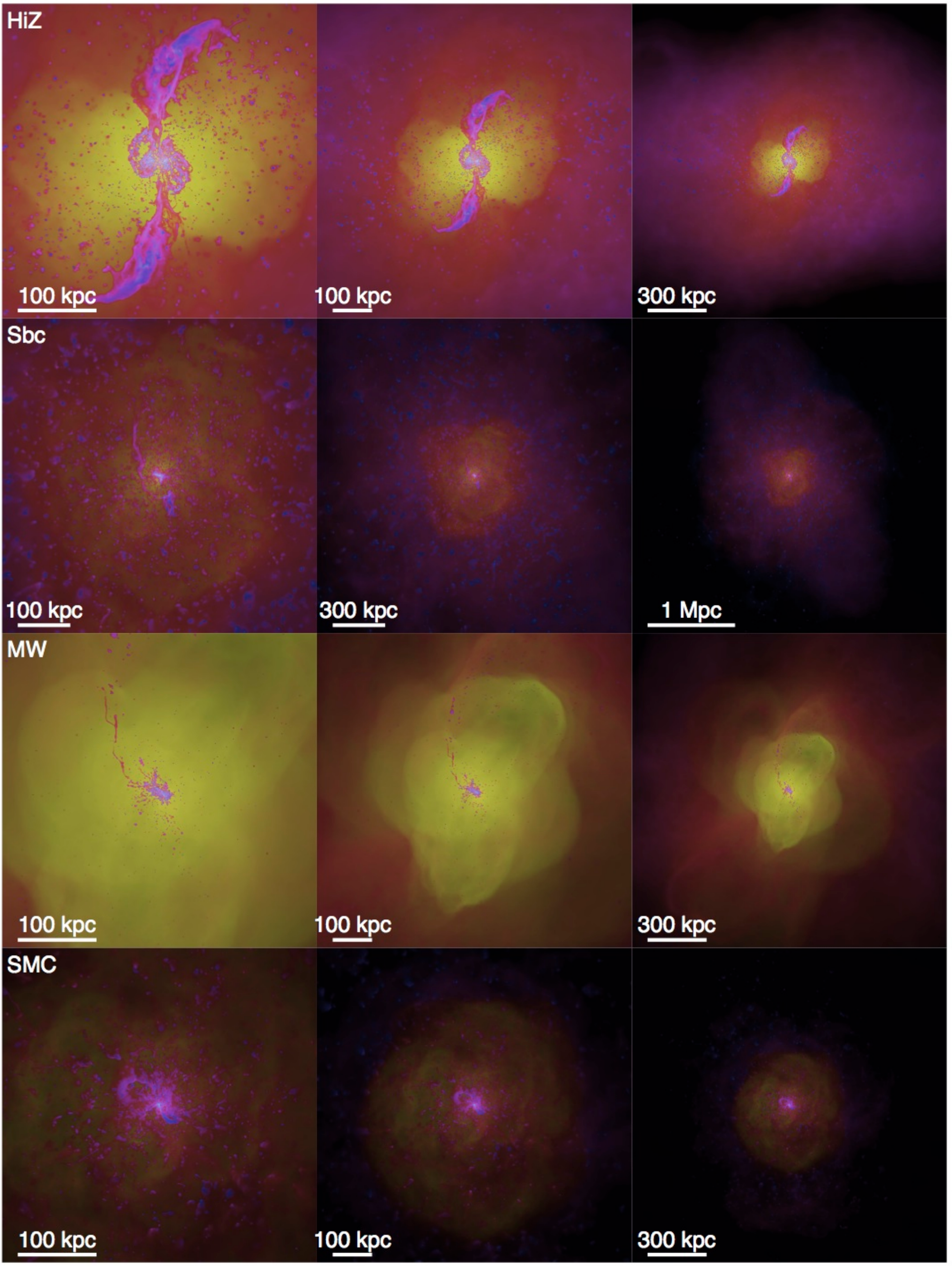}{0.7}
    \caption{Images of the gas in the starburst-driven superwinds in our simulations. Brightness encodes projected gas density (increasing with density; logarithmically scaled with a $\approx4\,$dex stretch); color encodes gas temperature with the blue/white material being $T\lesssim1000\,$K cold atomic/molecular gas, pink $\sim10^{4}-10^{5}$\,K warm ionized gas, and yellow $\gtrsim10^{6}\,$K hot gas. We show just one example of each merger, but the {\bf e} and {\bf f} orbits are similar in each case. For each, we show the image at a fixed time near the starburst and at various spatial scales. A massive wind is plainly evident; the winds are multi-phase with volume-filling hot gas and ejected streams/shells of cold gas. We caution that the formation of the small, marginally-resolved isolated ``blobs'' of cold gas within the hot background outflow (not the warm/cold shells or GMCs within the disk) is quite sensitive to the numerical details of the simulations.$^{\ref{foot:clump.caveats}}$ A large fraction is unbound and escapes the galaxy halo. In the MW case, the gas-poor nature of the merger means the wind is almost entirely ``hot''; in the Sbc and HiZ cases a much larger fraction of ejected material is warm/cool; in the SMC case there is a broad mix of warm gas and hot gas ejected. The shells and features in the diffuse gas arise from multiple bursty episodes shocking as they ``catch up'' to one another. 
    \label{fig:wind.images}}
\end{figure*}

\vspace{-0.5cm}
\section{Methods}
\label{sec:sims}

The simulations analyzed in this paper are presented in \paperfour. We therefore only briefly summarize their most important properties here, and refer interested readers to that paper for the simulation details. 

The simulations follow the methodology and galaxy models originally presented in \paperone\ (Sec.~2 \&\ Tables~1-3) and \papertwo\ (Sec.~2), using a heavily modified version of the TreeSPH code {\small GADGET-3} \citep{springel:gadget}, in its fully conservative ``density-entropy'' form \citep{springel:entropy}. They include stars, dark matter, and gas (with cooling, star formation, and stellar feedback). 

\vspace{-0.5cm}
\subsection{Initial Conditions}

We consider mergers of four initial disk models spanning a range of galaxy types. Each has a bulge, stellar and gas disk, halo, and central BH.\footnote{To isolate the effects of stellar feedback, models for BH growth and feedback are disabled here.} At our standard resolution, each model has $\approx 0.3-1\times10^{8}$ total particles, giving particle masses of $500-1000\,\msun$ and $1-5$\,pc smoothing lengths.\footnote{These are typical smoothing lengths in the dense gas; generally the smoothing lengths evolve adaptively following \citet{springel:entropy} to enclose a fixed number $\approx128$ neighbors. The gravitational softening lengths are set to be approximately equal to the minimum smoothing lengths.} Convergence tests of isolated versions of these disks have been extended to $\approx 10^{9}$ particles and sub-pc resolution.\footnote{These tests are described in \paperone\ and \papertwo, and used to check convergence in small-scale ISM properties. We have also run every simulation described in this paper with $10$ times fewer particles ($2$ times larger softening/smoothing); although some small-scale properties differ, our conclusions regarding quantities considered in this paper are identical. Additional numerical tests of the SPH method, relevant primarily for the wind phase structure in the extended halo, are presented in Appendix~\ref{sec:appendix:altsph}.} 

The disk models include: 

(1) SMC: an SMC-like dwarf, with baryonic mass $M_{\rm bar}=8.9\times10^{8}\,\msun$ 
and halo mass $M_{\rm halo}=2\times10^{10}\,\msun$ (concentration $c=15$), 
a \citet{hernquist:profile} profile bulge with a mass $m_{b}=10^{7}\,\msun$, and exponential 
stellar ($m_{d}=1.3\times10^{8}\,\msun$) and gas disks ($m_{g}=7.5\times10^{8}\,\msun$) 
with scale-lengths $h_{d}=0.7$ and $h_{g}=2.1$\,kpc, respectively. 
The initial stellar scale-height is $z_{0}=140$\,pc and the stellar disk is initialized such that the 
Toomre $Q=1$ everywhere.\footnote{The scale height is set to be $=0.2$ times the scale-length. Since the initial mass of stars is small this has little effect, and most new stars form with a somewhat larger scale-height; see \papertwo\ (Fig.~9) for further discussion.} The gas and stars are initialized with uniform metallicity $Z=0.1\,Z_{\sun}$.

(2) MW: a MW-like galaxy, with halo and baryonic properties of $(M_{\rm halo},\,c)=(1.6\times10^{12}\,\msun,\,12)$ and $(M_{\rm bar},\,m_{b},\,m_{d},\,m_{g})=(7.1,\,1.5,\,4.7,\,0.9)\times10^{10}\,\msun$, $Z=Z_{\sun}$, and scale-lengths/heights $(h_{d},\,h_{g},\,z_{0})=(3.0,\,6.0,\,0.3)\,{\rm kpc}$ (note the gas disk is more extended than the stellar disk, giving a gas fraction $\approx10\%$ inside the solar circle).

(3) Sbc: a LIRG-like galaxy (i.e.\ a more gas-rich spiral than is characteristic 
of those observed at low redshifts)
with $(M_{\rm halo},\,c)=(1.5\times10^{11}\,\msun,\,11)$,  
$(M_{\rm bar},\,m_{b},\,m_{d},\,m_{g})=(10.5,\,1.0,\,4.0,\,5.5)\times10^{9}\,\msun$, 
$Z=0.3\,Z_{\sun}$, 
and $(h_{d},\,h_{g},\,z_{0})=(1.3,\,2.6,\,0.13)\,{\rm kpc}$. 

(4) HiZ: a high-redshift massive starburst disk, chosen to match the 
properties of the observed non-merging but rapidly star-forming SMG 
population, with 
$(M_{\rm halo},\,c)=(1.4\times10^{12}\,\msun,\,3.5)$ and a 
virial radius appropriately rescaled for a halo at $z=2$ rather than $z=0$,  $(M_{\rm bar},\,m_{b},\,m_{d},\,m_{g})=(10.7,\,0.7,\,3,\,7)\times10^{10}\,\msun$, $Z=0.5\,Z_{\sun}$, and 
$(h_{d},\,h_{g},\,z_{0})=(1.6,\,3.2,\,0.32)\,{\rm kpc}$. 

We consider equal-mass mergers of identical copies of galaxies (1)-(4), on parabolic orbits with two representative choices for the initial disk orientations. The first (orbit {\bf e} in \citealt{cox:kinematics}) is near-prograde (a strong resonant interaction), the second (orbit {\bf f}) is near-retrograde (or polar-retrograde, a weak out-of-resonance interaction). For the most relevant properties of stellar winds, there is little difference between these orbits which bracket the range from most to least violent encounters; we therefore expect the properties examined here to be robust over a wide range of configurations.

\vspace{-0.5cm}
\subsection{Cooling \&\ Feedback}
\label{sec:sims:feedback}

Gas follows an atomic cooling curve with additional fine-structure cooling to $\sim10\,$K. Star formation is allowed only in dense, molecular, self-gravitating regions above $n>1000\,{\rm cm^{-3}}$. We follow \citet{krumholz:2011.molecular.prescription} to calculate the molecular fraction in dense gas as a function of local column density and metallicity, and follow \citet{hopkins:binding.sf.prescription} to calculate the local virial parameter of the gas (in order to restrict star formation to gas which is locally self-gravitating). This then forms stars at a rate $\dot{\rho}_{\ast}=\rho_{\rm mol}/t_{\rm ff}$; however the average efficiency on larger scales is much lower because of feedback.\footnote{In \paperone\ and \papertwo\ we show that the galaxy structure and SFR are basically independent of the small-scale SF law, because they are feedback-regulated. For example, we have re-run lower-resolution tests with a simpler prescription where star formation is restricted to all gas with $n>1000\,{\rm cm^{-3}}$, and find it makes little difference except to ``smear out'' the SFR in dense regions. As a result, this choice also has little effect on the winds studied here.}

Once stars form, their feedback effects are included from several sources:

(1) {\bf Momentum Flux} from Radiation Pressure, Supernovae, \&\ Stellar Winds: Gas surrounding stars receives a direct momentum flux $\dot{P}=\dot{P}_{\rm SNe}+\dot{P}_{\rm w}+\dot{P}_{\rm rad}$, where the separate terms represent the direct momentum flux of SNe ejecta, stellar winds, and radiation pressure. The first two are directly tabulated for a single stellar population as a function of age and metallicity $Z$ and the flux is directed away from the star. The latter is approximately $\dot{P}_{\rm rad}\approx (1+\tau_{\rm IR})\,L_{\rm incident}/c$, where $1+\tau_{\rm IR} = 1+\Sigma_{\rm gas}\,\kappa_{\rm IR}$ accounts for the absorption of the initial UV/optical flux and multiple scatterings of the IR flux if the region between star and gas particle is optically thick in the IR.\footnote{Photons which escape the local stellar vicinity can be absorbed at larger radii. Knowing the intrinsic spectrum of each star particle, we attenuate integrating the local gas density and gradients to convergence, and propagate the resulting ``escaped'' flux to large distances. This can then be used to calculate the local incident flux on all gas particles, from which local absorption is calculated by integrating over a frequency-dependent opacity that scales with metallicity. The appropriate radiation pressure force is then imparted.}

(2) {\bf Supernova Ejecta \&\ Shock-Heating}: Gas shocked by supernovae can be heated to high temperatures. We tabulate the SNe Type-I and Type-II rates from \citet{mannucci:2006.snIa.rates} and STARBURST99, respectively, as a function of age and metallicity for all star particles and stochastically determine at each timestep if a SNe occurs. If so, the appropriate mechanical luminosity is injected as thermal energy in the gas within a smoothing length of the star particle, along with the relevant mass and metal yield. 

(3) {\bf Gas Recycling and Shock-Heating in Stellar Winds:} Similarly, stellar winds are assumed to shock locally and so we inject the appropriate tabulated mechanical power $L(t,\,Z)$, mass, and metal yields, as a continuous function of age and metallicity into the gas within a smoothing length of the star particles. The integrated mass fraction recycled is $\sim0.3$.  

(4) {\bf Photo-Heating of HII Regions and Photo-Electric Heating}: We also tabulate the rate of production of ionizing photons for each star particle; moving radially outwards from the star, we then ionize each neutral gas particle until the photon budget is exhausted. Ionized gas is maintained at a minimum $\sim10^{4}\,$K until it falls outside an HII region. Photo-electric heating is followed in a similar manner using the heating rates from \citet{wolfire:1995.neutral.ism.phases}.

Extensive numerical tests of the feedback models are presented in \papertwo. All energy, mass, and momentum-injection rates are taken from the stellar population models in STARBURST99, assuming a \citet{kroupa:imf} IMF, without free parameters.

\begin{figure*}
    \centering
    \plotsidesize{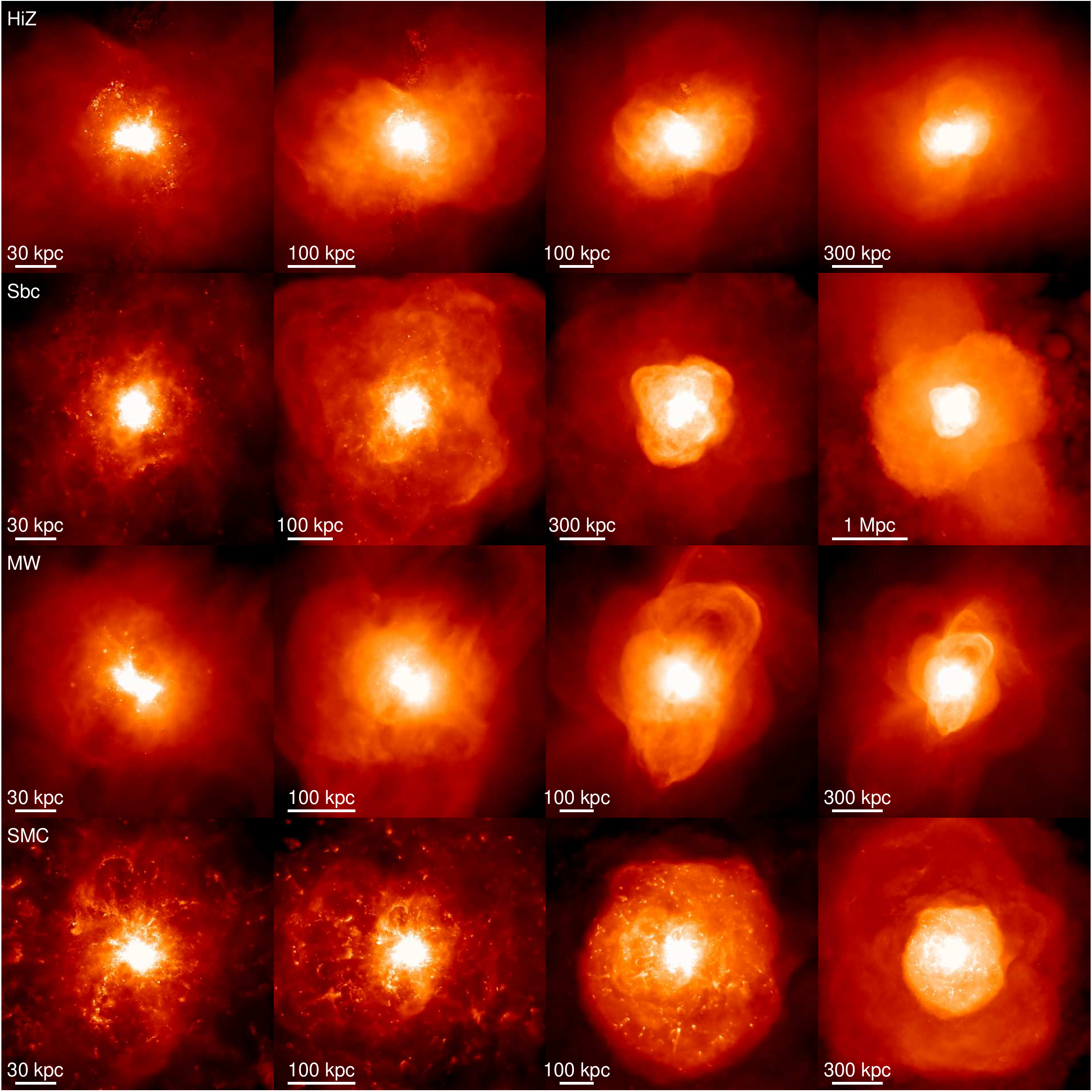}{1.0}
    \caption{Galactic wind thermal+metal line emission morphology, as a proxy for X-ray emission (though we caution this includes gas with a broad range of temperatures). The maps show the images as Fig.~\ref{fig:wind.images}, but in a single-color scale where intensity here encodes the projected bremsstrahlung emissivity plus metal cooling luminosity (again we caution that the small clumps at large radii may be artificial$^{\ref{foot:clump.caveats}}$).  Since this weights the volume-filling hot gas, the different bubbles and shells are more clear. At each scale, clear changes in the orientation of the older/larger bubbles are also evident.
    \label{fig:wind.images.xr}}
\end{figure*}

\vspace{-0.5cm}
\section{Outflow Morphologies}
\label{sec:outflow.morph}

For reference, Figure~\ref{fig:morph.demo.stars} shows the stellar morphology as it would be optically observed with ideal resolution during a representative stage of the merger simulations, when all feedback mechanisms are present (the image is a mock $u/g/r$ composite calculated as described in \paperfour). 

In this paper, we are interested in the structure of the outflows during the mergers. This is shown in Fig.~\ref{fig:wind.images}. Here we show the projected gas density with colors encoding the gas temperature. The projected temperatures are logarithmically-averaged and surface-density weighted, so reflect the temperature of most of the line-of-sight gas mass, rather than the temperature that contains most of the thermal energy.

Outflows are plainly evident; we will quantify their properties and phase distribution in detail below. Briefly, on small scales the simulated ISM is a supersonically turbulent medium \citep[see][]{hopkins:excursion.ism,hopkins:excursion.imf} in which cold giant molecular clouds (GMCs) continuously form and are dispersed by feedback after turning a few percent of their mass into stars (see \papertwo, \paperfour). The sub-galactic structure is qualitatively similar to that seen in other simulations with similar resolution and explicit treatment of the cold gas \citep[e.g.][]{bournaud10}, albeit with some significant differences owing to which feedback mechanisms are or are not included (for a detailed comparison, see \paperfour). There is volume-filling hot gas (heated by SNe and O-star winds), which vents an outflow component; occasionally the early stages are evident as ``bubbles'' breaking out of the warm phase gas. Warm/cold gas is mixed throughout the outflow,\footnote{\label{foot:clump.caveats}We caution that the very small, cold ``clumps'' at large radii in the wind (distinct from the resolved molecular clouds in the disk and the coherent warm/cold shells and filaments in outflow) are only marginally resolved and are probably an artifact of the numerical method used, in concert with the fact that there is no cosmic ionizing background present in these simulations to heat gas far from the galaxy. In Appendix~\ref{sec:appendix:altsph} we re-run a limited subset of our simulations with an alternative formulation of SPH designed to treat contact discontinuities more accurately, and include UV background heating, and show that these largely disappear. We will focus in this paper only on results robust to our resolution and numerical method, but direct the interested reader to Appendix~\ref{sec:appendix:altsph} and \citet{hopkins:lagrangian.pressure.sph} for more details.} especially near the galaxy (before they have had time to mix with the more diffuse material); these are sometimes entrained by the hot gas but more often directly accelerated by radiation pressure. Qualitatively, this is true in isolated galaxies as well (\paperthree). 

In Figure~\ref{fig:wind.images}, successive generations of ``bursts'' and strong outflows are evident as overlapping shells (many caused by shocks as different ejection events ``catch up''); these are each broadly associated with a galaxy passage. The diffuse hot gas, being volume-filling, has a near-unity covering factor, but it is still clearly organized into various ``bubbles.'' Each of these has a broadly bimodal morphology set by much of the hot gas blowing out perpendicular to the disk along the path of least resistance. Since the disk orientations change during the merger owing to gravitational torques, the successive bursts have different orientations. At the largest radii, the diffuse gas cools adiabatically; this would not necessarily occur with a realistic IGM present into which the hot gas could propagate. The warm and cool gas also has a significant (albeit smaller) covering factor even at $\gtrsim100\,$kpc (except in the MW model). However, the morphology is much less smooth -- this gas is primarily in filaments, and shells. Although the crude average distribution of these is similar to the hot gas, there is much larger line-of-sight variation. Some (but not all) of this is material is accelerated by radiation pressure, and so tends to reach somewhat lower velocities than the hottest pressure-accelerated gas, and therefore the density falls off proportionally more rapidly at very large radii (although, being more dense, this material may be able to propagate ballistically into the IGM where lower-density, volume-filling material would be halted by the ambient pressure). In any case, as we discuss in detail in \paperthree\ and \S~\ref{sec:discussion}, the outflows should not be taken too literally at the largest radii: we do not include initial gaseous halos or a realistic IGM into which the wind should propagate, so the winds here expand unimpeded beyond the halo. 

X-ray observations provide a strong probe of the hot phase of the galactic winds; Figure~\ref{fig:wind.images.xr} therefore shows the same images, but now in their approximate X-ray properties. For convenience, rather than make a detailed mock observation corresponding to a given instrument, sensitivity, redshift, and energy range, we instead quantify the approximate X-ray emission with the sum of the thermal bremsstrahlung emission (emissivity per unit volume $u_{X} \propto T_{\rm gas}^{1/2}\,n_{e}\,n_{i}$, in terms of the gas temperature $T_{\rm gas}$, and electron/ion number densities $n_{e}$/$n_{i}$), and the metal-cooling luminosity (using the compiled tables in \citealt{wiersma:2009.coolingtables} as a function of $n_{i}$, $n_{e}$, $Z$, and $T_{\rm gas}$, assuming solar abundance ratios and the $z=0$ ionizing background). Note that these can include significant contributions from low-temperature gas so this need not refer specifically to X-ray observations. Broadly, the morphology is similar, but with the hot, low-density material highlighted, the various bubbles and shells are more obvious. It is also clear, as we go to larger scales, that the larger/older bubbles have distinct orientations, corresponding to the orientation of the galaxies at earlier merger stages.

\begin{figure*}
    \centering
    \plotsidesize{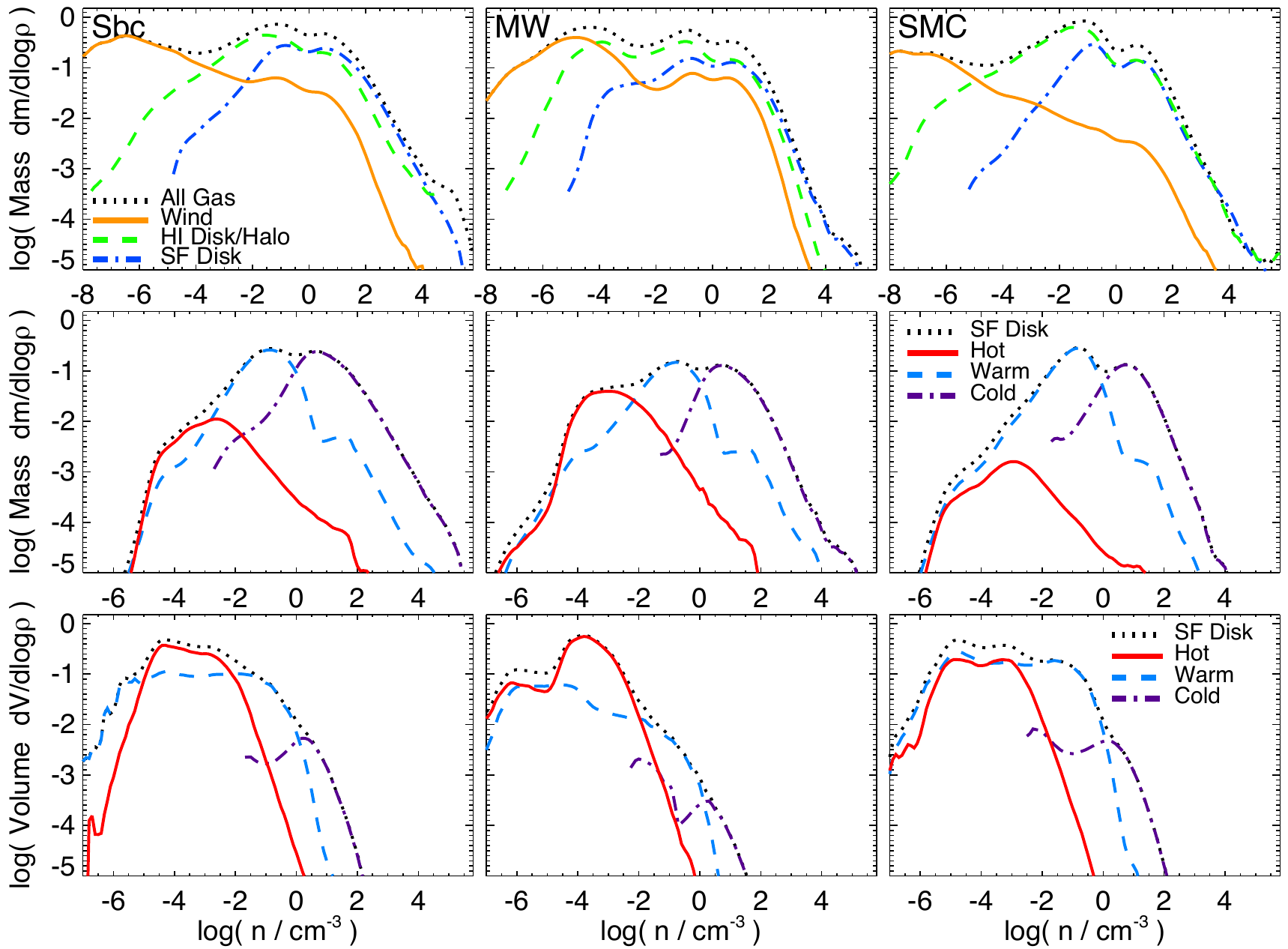}{1.0}
    \caption{Density distribution of different ISM phases, 
    for different galaxy disk models (Sbc, MW, and SMC), averaged over the duration of the merger. {\em Top:} 
    Mass-weighted density PDF (${\rm d}\,m_{\rm gas}/{\rm d}\log{n}$), 
    i.e.\ the mass fraction per logarithmic interval in density $n$. 
    We show the distribution for all gas in the simulation 
    ({\em black}), the gas approximately within the (multi-phase) star-forming disks ({\em blue}), 
    the gas in the extended, ionized disks and halo ({\em green}), 
    and wind/outflow material ({\em orange}). 
    These trace the material at high, intermediate, and low densities, 
    respectively (as expected). Each density PDF  has a very broad density distribution. 
    {\em Middle:} 
    Mass-weighted density PDF within the ``star-forming disks''. 
    We show the density PDF for all of the gas in the region 
    ({\em black}), the ``cold'' phase ({\em purple}; $T<2000$\,K), 
    the ``warm ionized'' phase ({\em cyan}; $2000<T<4\times10^{5}\,K$), 
    and the ``hot diffuse'' phase ({\em red}; $T>4\times10^{5}\,K$). 
    Most of the mass is in the cold phase (which 
    dominates at high densities), but with a comparable contribution 
    from the warm medium (the hot phase contributing a few percent). 
    The multi-modal nature of the total density PDF is a consequence of the strong phase separation. 
    Note that since these simulations do not include the cosmic ionizing background, 
    we truncate the cold-phase distribution at densities below which post-processing calculations suggest it would be photo-ionized ($\sim0.01\,{\rm cm^{-3}}$; see \citealt{faucher:ion.background.evol}); such gas contributes negligibly to the total at very low densities, in any case. 
    {\em Bottom:} Volume-weighted 
    density PDF (${\rm d}\,V_{\rm gas}/{\rm d}\log{n}$) for  the star forming disk. 
    The hot diffuse phase dominates, with a moderate 
    volume filling fraction for the warm phase, 
    and a small ($\sim1-5\%$) volume filling fraction of cold molecular clouds in the disk. 
    \label{fig:rho.dist}}
\end{figure*}

\begin{figure}
    \centering
    \plotonesize{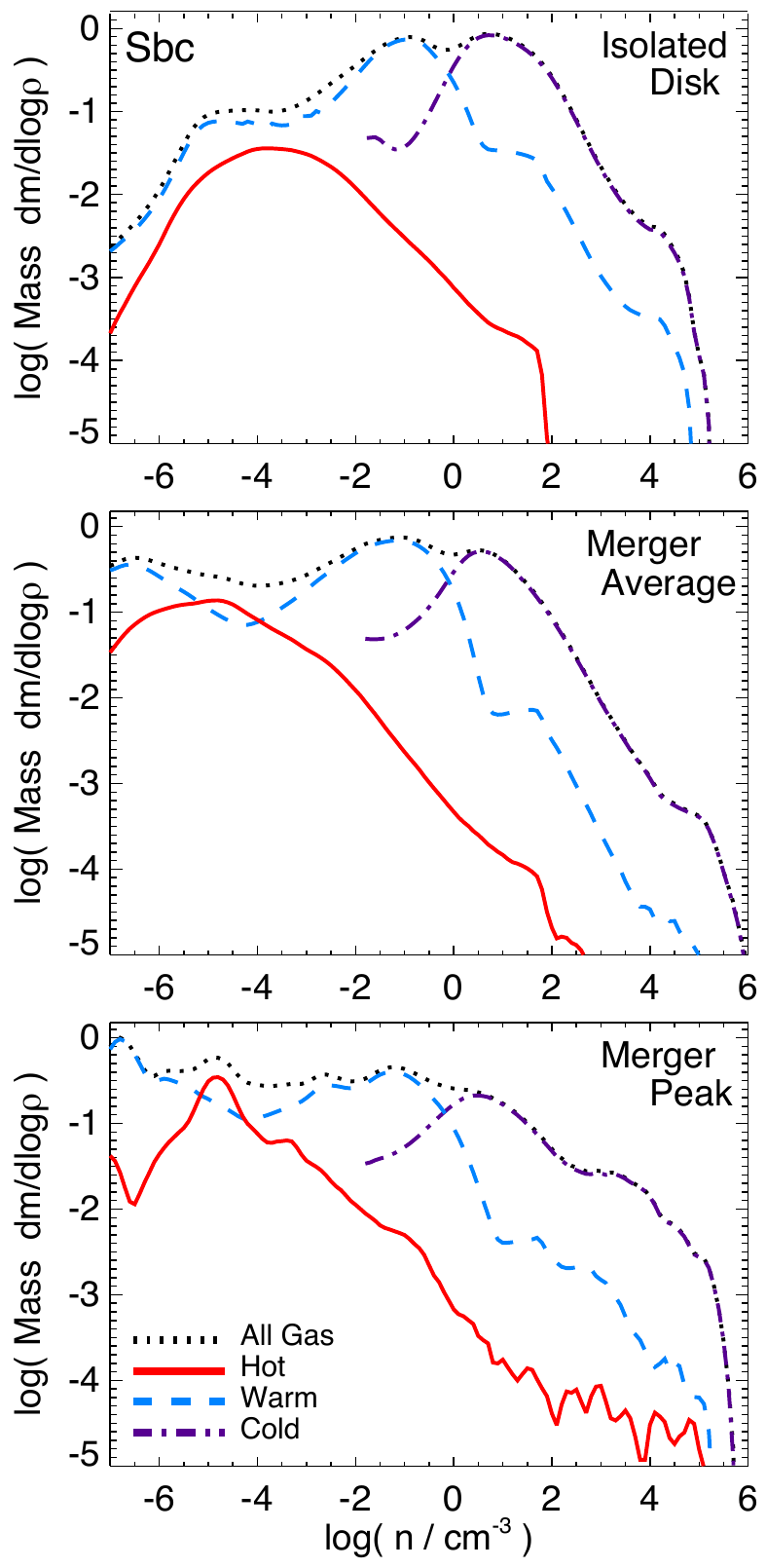}{0.8}
    \caption{Density distribution of different ISM phases (as Fig.~\ref{fig:rho.dist}), for all the gas in the simulation in one example (the Sbc {\bf e} case, but the results are qualitatively similar for each model). {\em Top:} The average, steady-state distribution for the isolated disk (see \papertwo). {\em Middle:} Distribution averaged over the duration of the merger. {\em Bottom:} Distribution at the snapshot where the merger-induced SFR is maximized. The average phase distribution over the entire merger is similar to (just slightly more broad than) the isolated disk counterpart, reflecting the fact that most of the SF is in the separate ``quiescent mode'' in the disks rather than the merger-induced burst. Unlike a no-feedback model, there is {\em not} a runaway pileup of gas at the highest densities. During the peak of activity, more gas is channeled to high densities $\gtrsim10^{4}\,{\rm cm^{-3}}$, but otherwise the distributions (and characteristic density of each temperature phase) are similar. This would be observable in e.g.\ the ratio of $L_{\rm HCN}/L_{\rm CO(1-0)}$ which would increase from $\approx 0.02$ in the isolated case (see \citealt{hopkins:dense.gas.tracers}) to $\approx 0.2-0.3$ in the peak of the merger.
    \label{fig:rho.dist.vstime}}
\end{figure}

\begin{figure*}
    \centering
    \plotsidesize{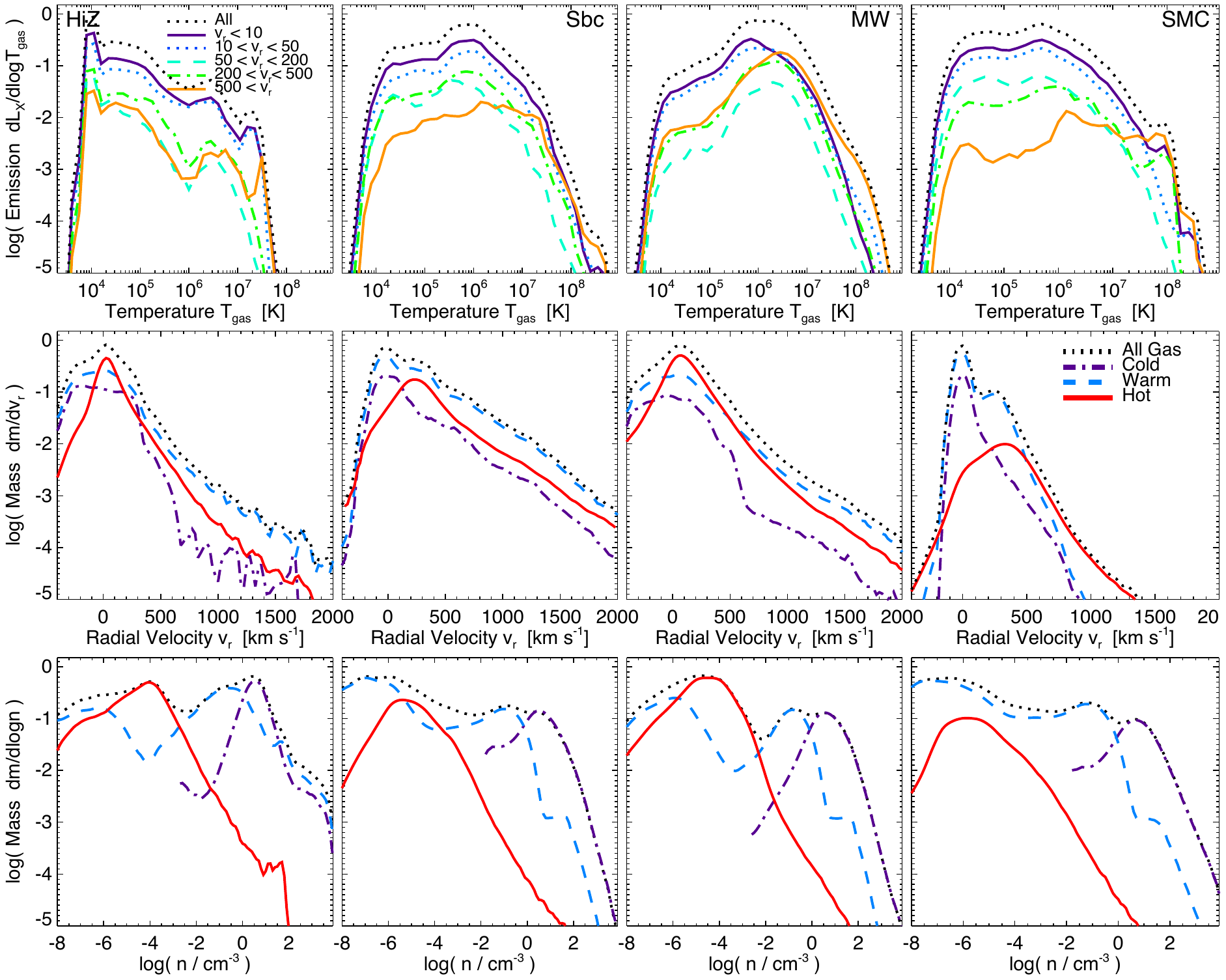}{1.}
    \caption{{\em Top:}  Distribution of gas temperature (weighted by thermal bremsstrahlung emission), 
    $dL_{X}/d\log{T_{\rm gas}}$, at $\approx200\,$Myr after the final coalescence in the merger.
    Different lines in each panel denote gas with different radial velocity (relative to the center-of-mass). 
    Emission from more rapidly-outflowing wind material is dominated by hotter gas ($T \sim 10^{6-8}$\,K). 
    {\em Middle:} 
    Mass-weighted distribution of gas outflow radial velocities ($dm/dv_{r}$).  
    We show separately the distribution from three phases: cold ($T<2000\,K$), 
    warm (primarily ionized) ($2000<T<4\times10^{5}\,K$) and hot/diffuse ($T>4\times10^{5}\,K$) gas.
    The outflows consist primarily of a mix of warm and hot gas, with some colder material. 
    Warm material typically dominates by mass, in the form of filaments and shells (in Fig.~\ref{fig:wind.images}; note the dense ``cloudlets'' at large radii, while visually prominent, are partly numerical and do not dominate by mass$^{\ref{foot:clump.caveats}}$).
    {\em Bottom:} 
    Mass-weighted density distribution for the wind gas, 
    divided into phases as in the middle panel.
    The ``warm'' material at very low densities, $\ll 10^{-4}\,{\rm cm^{-3}}$ 
    is previously ``hot'' material that has adiabatically cooled as it expands; 
    these low densities arise artificially because we do not include an IGM into which the wind expands.
    \label{fig:wind.phases}}
\end{figure*}

\vspace{-0.5cm}
\section{Phase Structure: Hot X-Ray Halos and Cold Molecular Gas}
\label{sec:wind.phases}

In \papertwo, we discuss in great detail the phase structure and density distribution of the ISM; here, we examine whether the same results obtain in mergers. Fig.~\ref{fig:rho.dist} plots the density PDF of the ISM, i.e.\ mass per logarithmic interval in $n$. For clarity, we just show the {\bf f} models, but the {\bf e} models are extremely similar. This covers a very wide dynamic range and is not directly comparable to the typical ``ISM density distribution'' from galaxy studies. To see this, we divide the gas into three categories: the ``star-forming disk(s)'' (gas within $R_{90}$, within one exponential ``scale height'' defined with respect to the angular momentum plane, and with outflow velocity $v_{r}<100\,{\rm km\,s^{-1}}$); the wind/outflow (defined below as un-bound gas with large outflow velocity), and extended disk+halo gas (the remaining gas; recall there is no initial extended gaseous halo in these simulations). Unsurprisingly, the SF ``disks'' include most of the dense gas, the winds include the least dense material (much of it out at or past the virial radius), and the ``halo'' is intermediate. We note again that since there is no IGM, escaped material can reach arbitrarily low densities. We also compare the distribution of density (weighted by mass or volume) in different temperature phases corresponding to cold/warm ionized/hot X-ray emitting gas within the star forming regions/disks (the exact temperature cuts are arbitrary but the qualitative comparison does not change if we shift them by moderate amounts). Here, we recover a qualitatively similar result to the isolated disks: the high-density gas ($>1\,{\rm cm^{-3}}$) is predominantly cold and contains most or $\sim$half the mass but has a small volume-filling factor of a couple percent; the intermediate-density gas ($0.01-1\,{\rm cm^{-3}}$) is primarily warm and has both a significant fraction of the mass ($\sim30-50\%$) and sizeable filling factors; the low-density gas ($<0.01\,{\rm cm^{-3}}$) is primarily hot and has order-unity volume filling factors. Each component is crudely (but not exactly) log-normal. As discussed in \papertwo, the turbulent pressure inside GMCs is much larger than the background pressure (they are marginally self-gravitating, rather than pressure-confined). 

Fig.~\ref{fig:rho.dist.vstime} shows how this compares (over the course of the merger) to the isolated disks. We show one example but the results are similar in each case. We plot the density distribution by temperature phase for all gas, averaged over the run of the isolated disk simulation (in which it reaches a steady-state, so is nearly time-independent), averaged over the merger duration, and at the snapshot with the peak SFR (near final coalescence) in the merger. Averaged over the merger, there is little difference (some material is at lower densities simply because the winds have more time to ``escape,'' and slightly more material is at high densities). This reflects the well-known fact that in merger simulations, most of the time (and most of the star formation) is contributed by the ``isolated mode,'' namely the separate SF in the two disks as they orbit between passages, rather than the merger-induced ``burst'' on top of this (which only dominates the central $\sim$kpc seen in \paperfour; see also \citealt{cox:massratio.starbursts,hopkins:sb.ir.lfs}). During the peak starburst, the results are still qualitatively similar (especially for the warm/hot gas); the peak density associated with each phase is also the same. The main difference is that a fraction of the gas funneled into the galaxy centers is pushed to much higher densities $n\gg10^{4}\,{\rm cm^{-3}}$ where it rapidly turns into stars. Although we stress that this is still not {\em most} of the gas (even the dense molecular gas). This is very different from models with weak/no feedback, which see catastrophic runaway to arbitrarily high densities $n\sim10^{6}$ for {\em most} of the star-forming gas, in contrast to observations which show that most of the gas in GMCs and other dense regions is at modest, non star-forming densities \citep[e.g.][and references therein]{williams:1997.gmc.prop,evans:1999.sf.gmc.review}.

The increase in gas at the highest densities during the starburst would be evident in dense molecular traces, as discussed in \citet{narayanan:co.outflows,hopkins:dense.gas.tracers}. Specifically, if we adopt the simple conversions therein from mass above $\sim10^{4}\,{\rm cm^{-3}}$ to HCN luminosity and mass above $\sim100\,{\rm cm^{-3}}$ to CO(1-0) luminosity, we estimate that the ratio $L_{\rm HCN}/L_{\rm CO(1-0)}$ should increase from $\approx 0.02$ in the isolated case (presented therein) to $\approx 0.15-0.30$ in the peak of the merger. This is consistent with what is seen in real local ULIRGs, essentially all of which are late-stage major mergers, in the compilation of \citet{gao:2004.hcn.compilation,narayanan:2008.sfr.densegas.corr,juneau:2009.enhanced.dense.gas.ulirgs}. 

In Fig.~\ref{fig:wind.phases} we show the phase distribution of the winds, similar to Fig.~\ref{fig:rho.dist}. These are discussed in detail for the isolated cases in \paperthree. We compare the distribution of temperatures weighted by their contribution to thermal bremsstrahlung plus metal cooling emission (as described above) for all the gas, in different outflow velocity intervals. We also plot the velocity distribution of all gas (mass per radial outflow velocity $v_{r}$); the ``peak'' near $v_{r}=0$ being the non-wind material (with the wind evident in the large tail). And we show the density distribution of the wind material specifically, in the style of Fig.~\ref{fig:rho.dist}. The velocity distributions are wide (discussed below). The inhomogeneous morphologies of the winds are reflected in the broad phase and density distribution (though recall the numerical caveats regarding the clumpy structure of the outflows). The extremely low-density wind material is a consequence of our not including a full IGM into which the winds can propagate; this also causes the winds to cool adiabatically (hence the secondary ``bump'' of warm/cold gas at extremely low densities). However it is fairly generic that the cold/warm/hot material dominates the wind at high/intermediate/low densities, respectively. It is especially worth noting that the winds can include some cold gas at densities $\sim 1-100\,{\rm cm^{-3}}$ (this is primarily material still near or within the disk, being accelerated outwards; as it escapes, the gas expands and can easily be heated).$^{\ref{foot:clump.caveats}}$ In the gas-rich mergers, the contribution to the wind mass from cold/warm/hot phases are comparable, although in the dwarf systems (SMC and Sbc) the very low-density ``warm'' component was mostly ``hot'' when originally ejected; in the gas-poor MW-like case, there is relatively little material to ``entrain'' and torques are able to efficiently force most of the dense gas into a starburst, so the outflow is much more strongly dominated by hot, venting gas.

\begin{figure*}
    \centering
    \scaleup
\begin{tabular}{ccc}
    \includegraphics[width={0.6\columnwidth}]{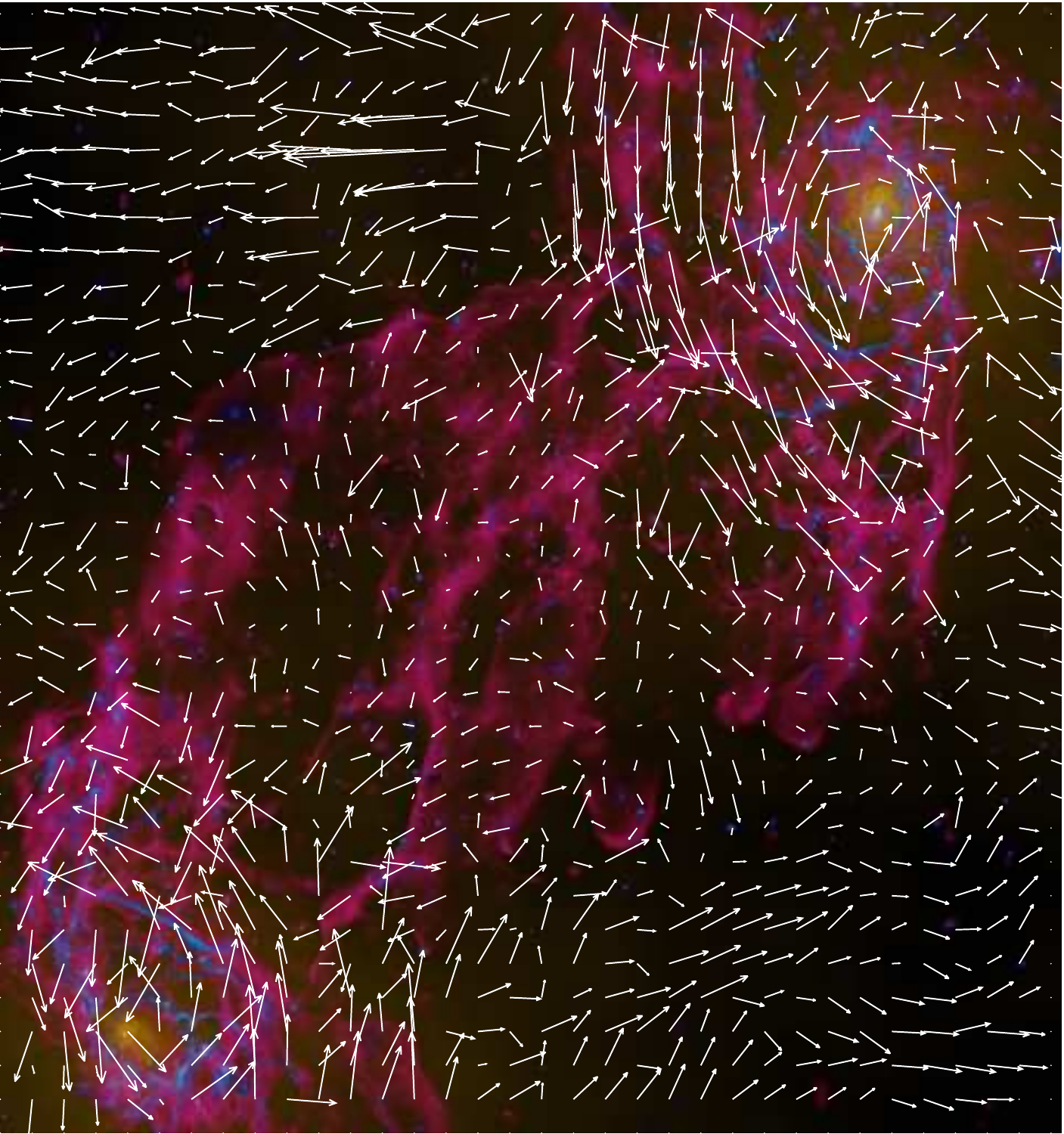} &
    \includegraphics[width={0.6\columnwidth}]{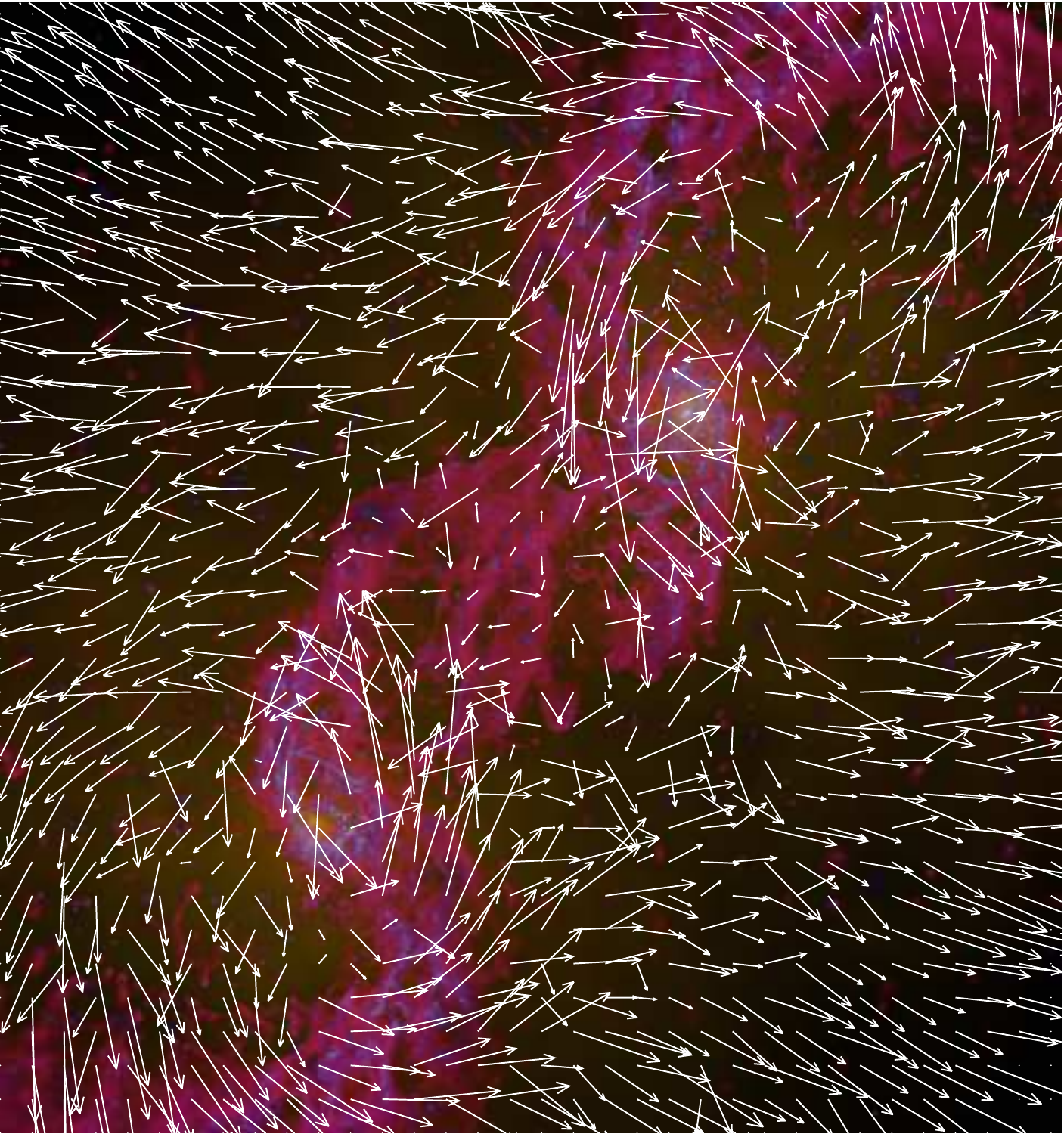} &
    \includegraphics[width={0.6\columnwidth}]{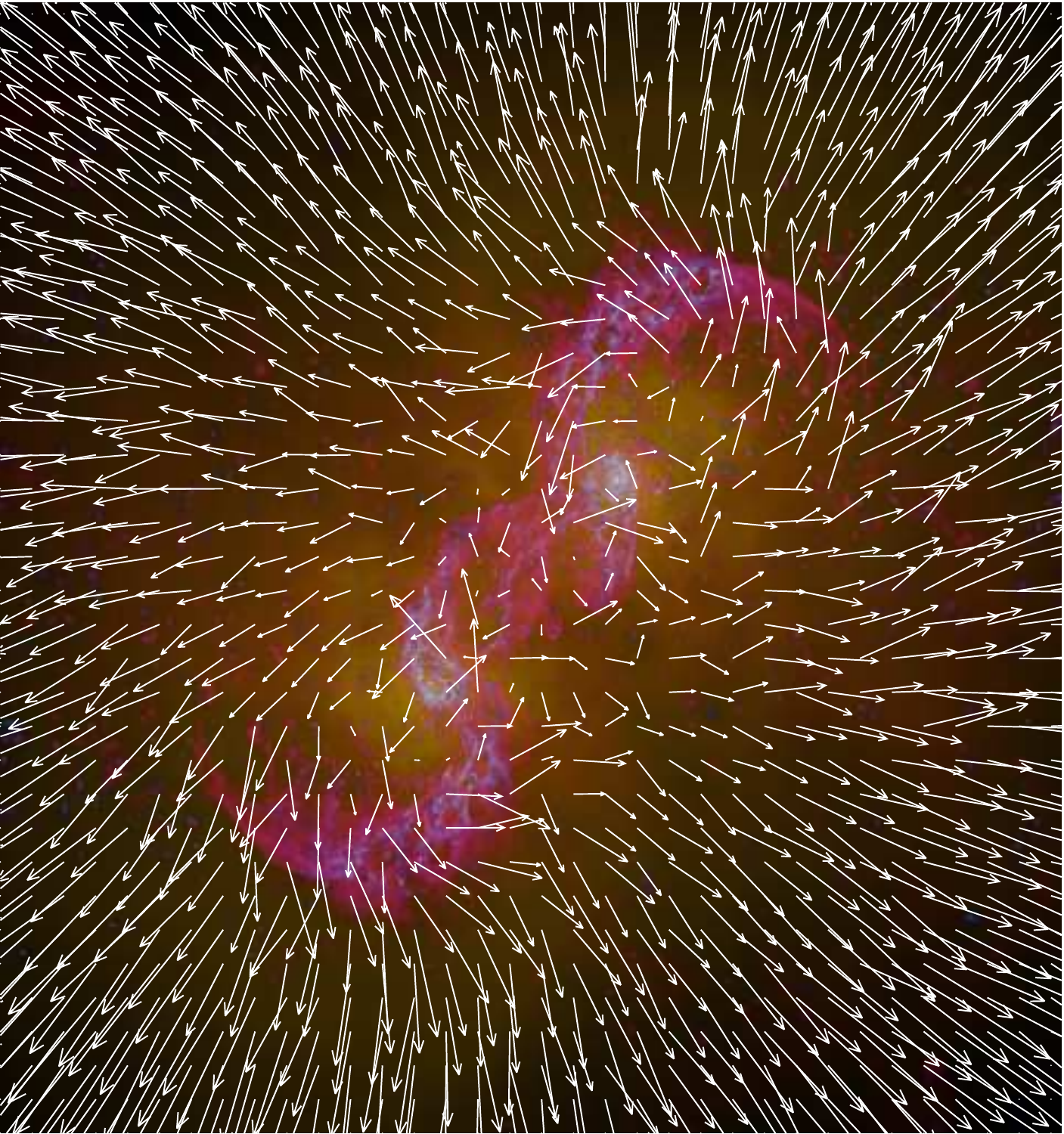} \\
    \includegraphics[width={0.6\columnwidth}]{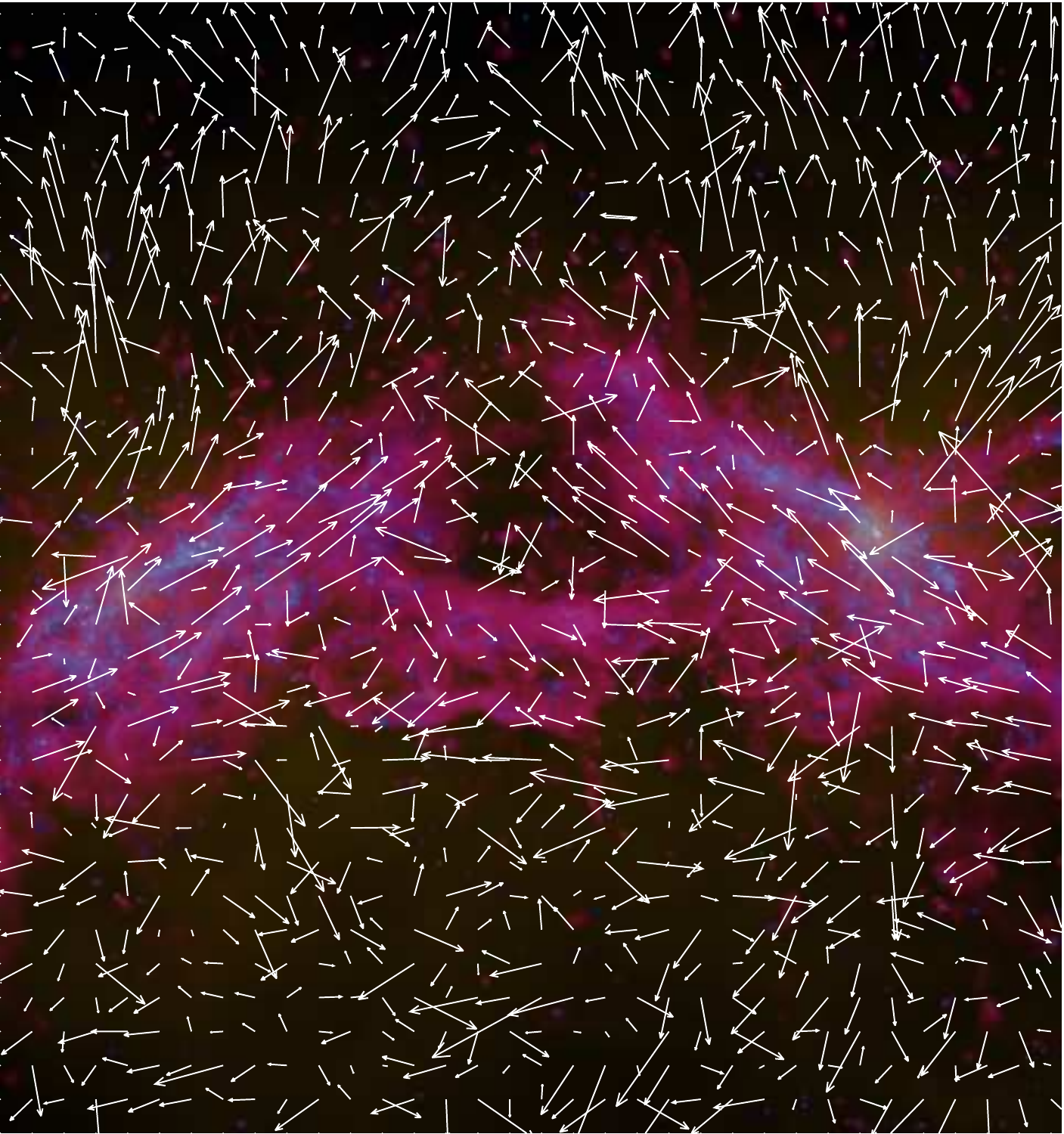} &
    \includegraphics[width={0.6\columnwidth}]{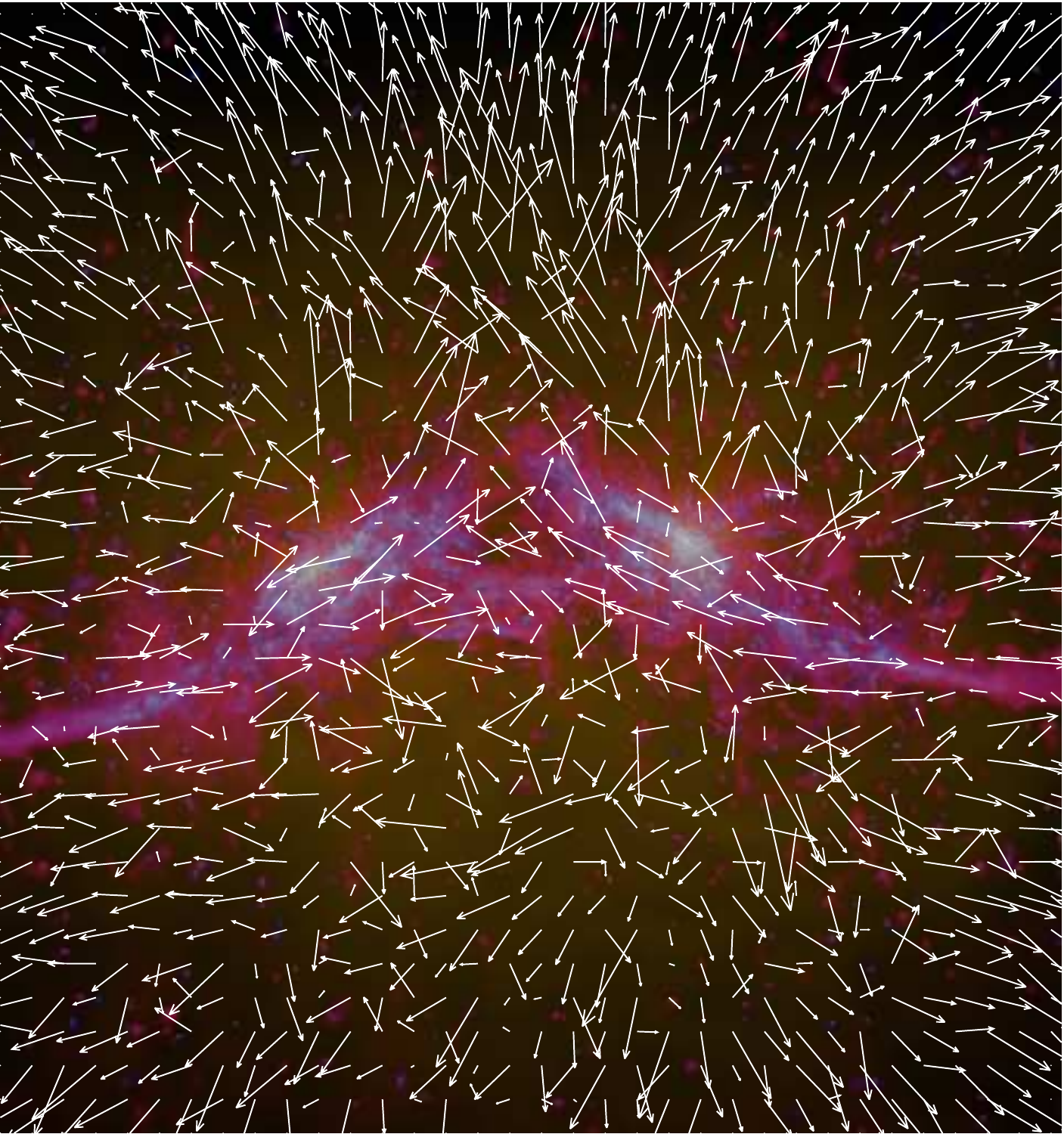} &
    \includegraphics[width={0.6\columnwidth}]{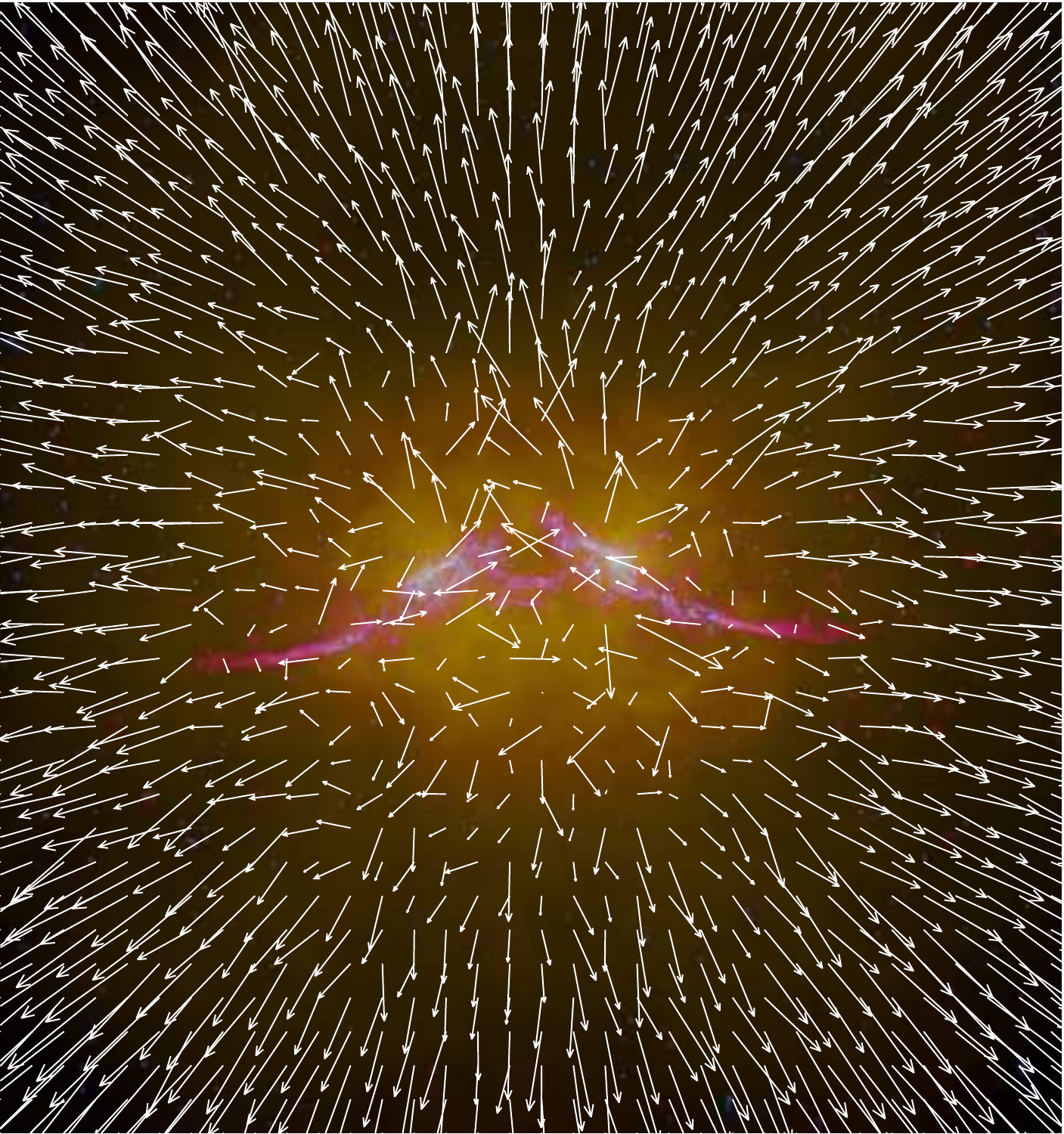} \\
\end{tabular}
    \caption{Image of the gas velocity field. The gas (color-coded in the style of Fig.~\ref{fig:wind.images}) is shown for the galaxy in Fig.~\ref{fig:morph.demo.stars} at the same time, in a face-on ({\em top}) and edge-on ({\em bottom}) projection, with the (projected in-plane) velocity vectors plotted. The vectors interpolate the gas velocities evenly over the image; their length is linearly proportional to the magnitude of the local velocity with the longest plotted corresponding to $\approx500\,{\rm km\,s^{-1}}$. The spatial scale of each image is $\pm50\,$kpc ({\em left}), $\pm100\,$kpc ({\em middle}), $\pm200\,$kpc ({\em right}). On scales near the galaxies the field is complex and the outflow ``launching'' region traces the entire disk surface, with significant non-radial components from the disk and merger orbital motion. On much larger scales the outflow is primarily radial.
    \label{fig:vfield}}
\end{figure*}

\begin{figure}
    \centering
    \plotonesize{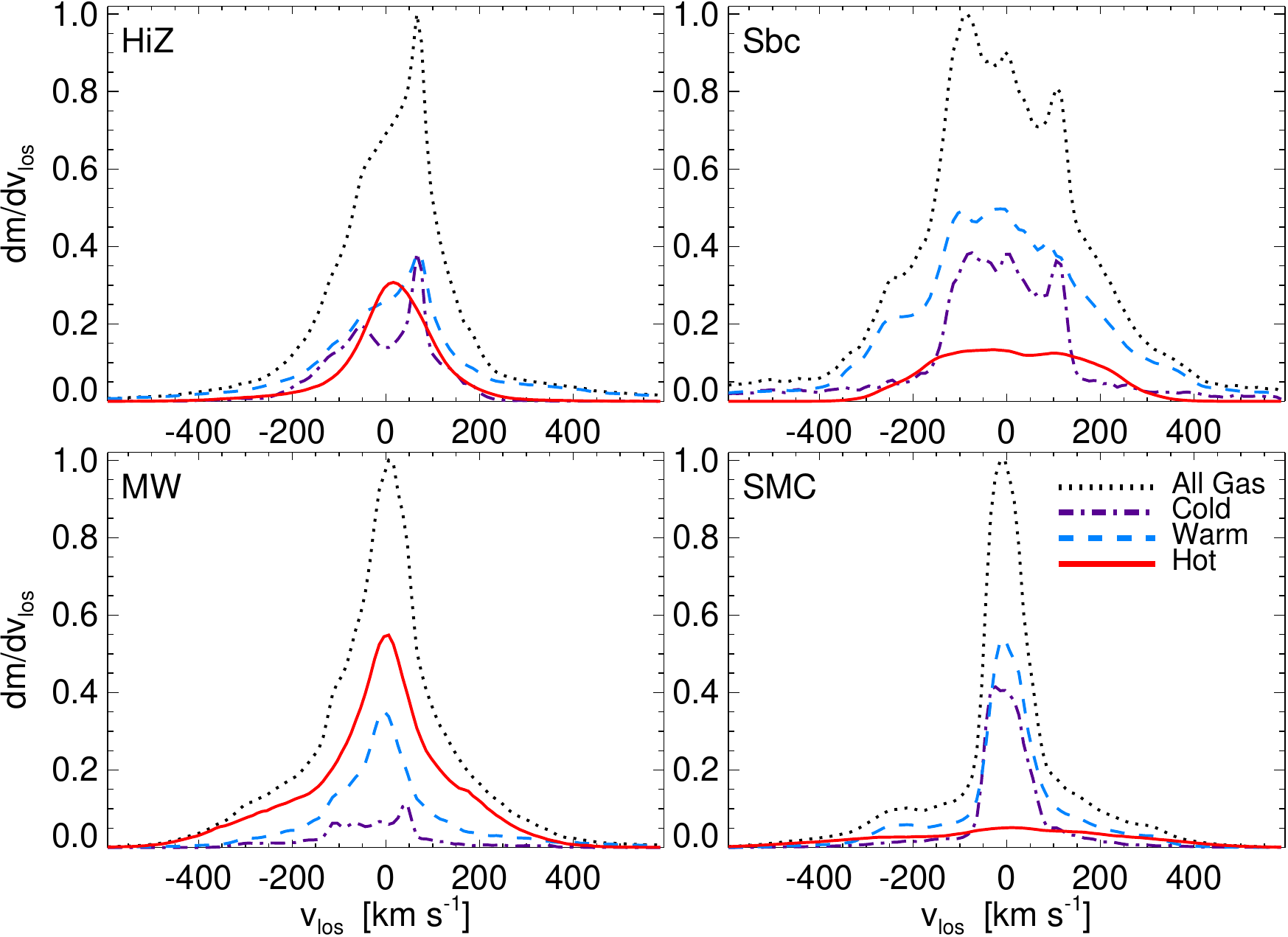}{1.0}
    \caption{Line-of-sight velocity distribution of the galaxies, on linear scale. Specifically, we project each galaxy along a random axis at a time near the peak starburst (just after final coalescence), and take the integrated ${\rm d}m/{\rm d}v_{\rm los}$ for all gas, and for the gas separated by temperature (as Fig.~\ref{fig:wind.phases}). Disturbed merger kinematics manifest as asymmetry in the ``core,'' the winds are evident in the broad ``wings'' extending to several hundred ${\rm km\,s^{-1}}$. We caution that this is not the same as an actual observed line profile (there is no accounting for emission/absorption here), but gives an idea of the line-of-sight kinematics in the relevant gas phases.
    \label{fig:integ.spec}}
\end{figure}

\begin{figure*}
    \centering
    \plotsidesize{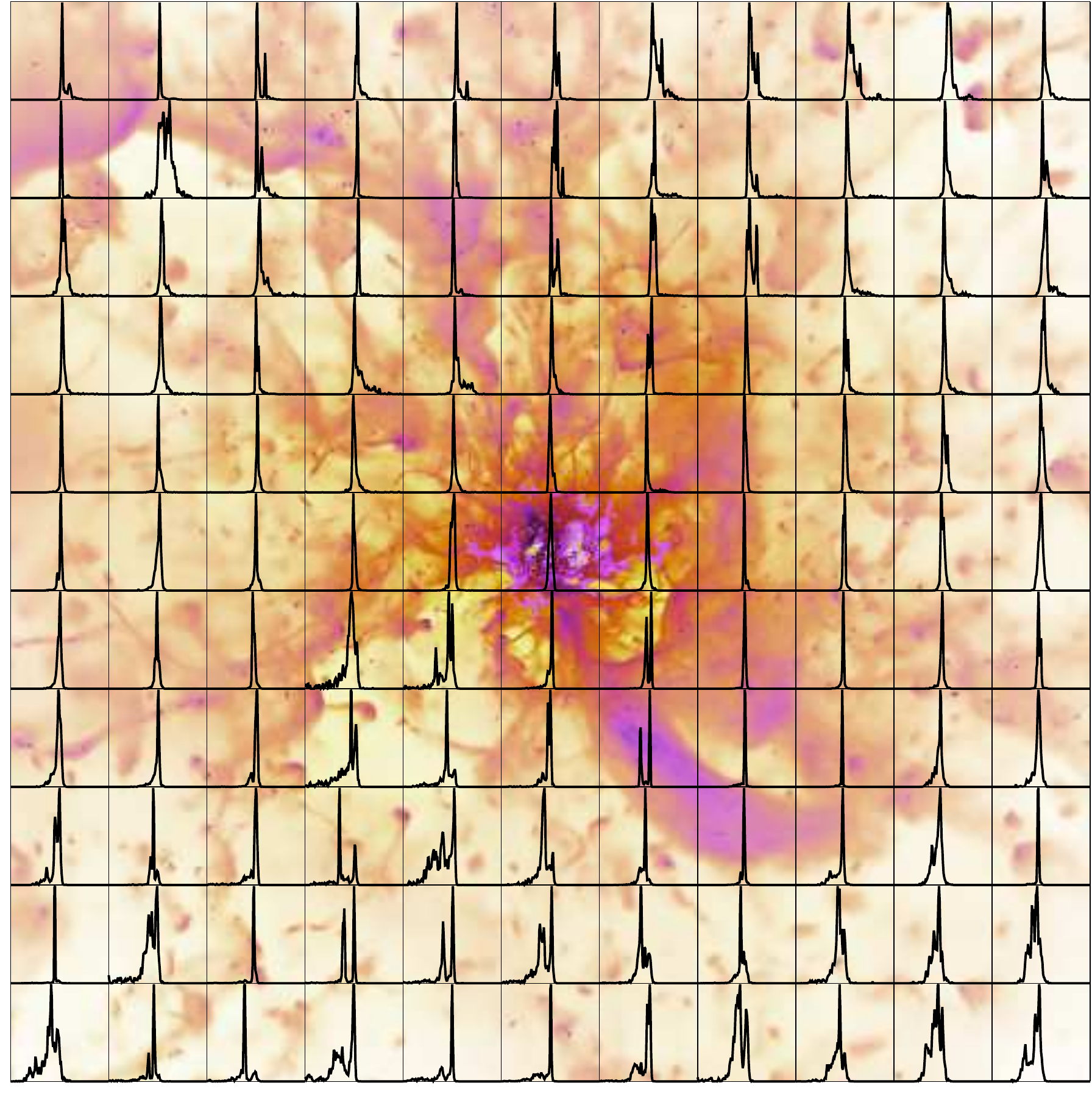}{1.0}
    \caption{Spatially resolved line-of-sight velocity distributions, for one example simulation (the SMC {\bf f} merger), at a time just after final coalescence. The background image is the simulation gas as Fig.~\ref{fig:wind.images}, but with inverted brightness (darker is more dense) for clarity. Each box overplots the projected LOSVD of all gas as in Fig.~\ref{fig:integ.spec}, averaged over a FWHM $5$\,kpc Gaussian aperture centered on the box center. The x-axis in each LOSVD box ranges from $\pm750\,{\rm km\,s^{-1}}$. Dense filaments appear as complex multiple narrow features (along with clumps, though we caution these may be numerical artifacts$^{\ref{foot:clump.caveats}}$); lines of sight through larger shells have a bimodal appearance; velocity centroid offsets are present at different radii but relatively small ($\sim100\,{\rm km\,s^{-1}}$). The broad velocity distribution is evident in the ubiquitous asymmetric tails. 
    \label{fig:ifu}}
\end{figure*}

\begin{figure}
    \centering
    \plotonesize{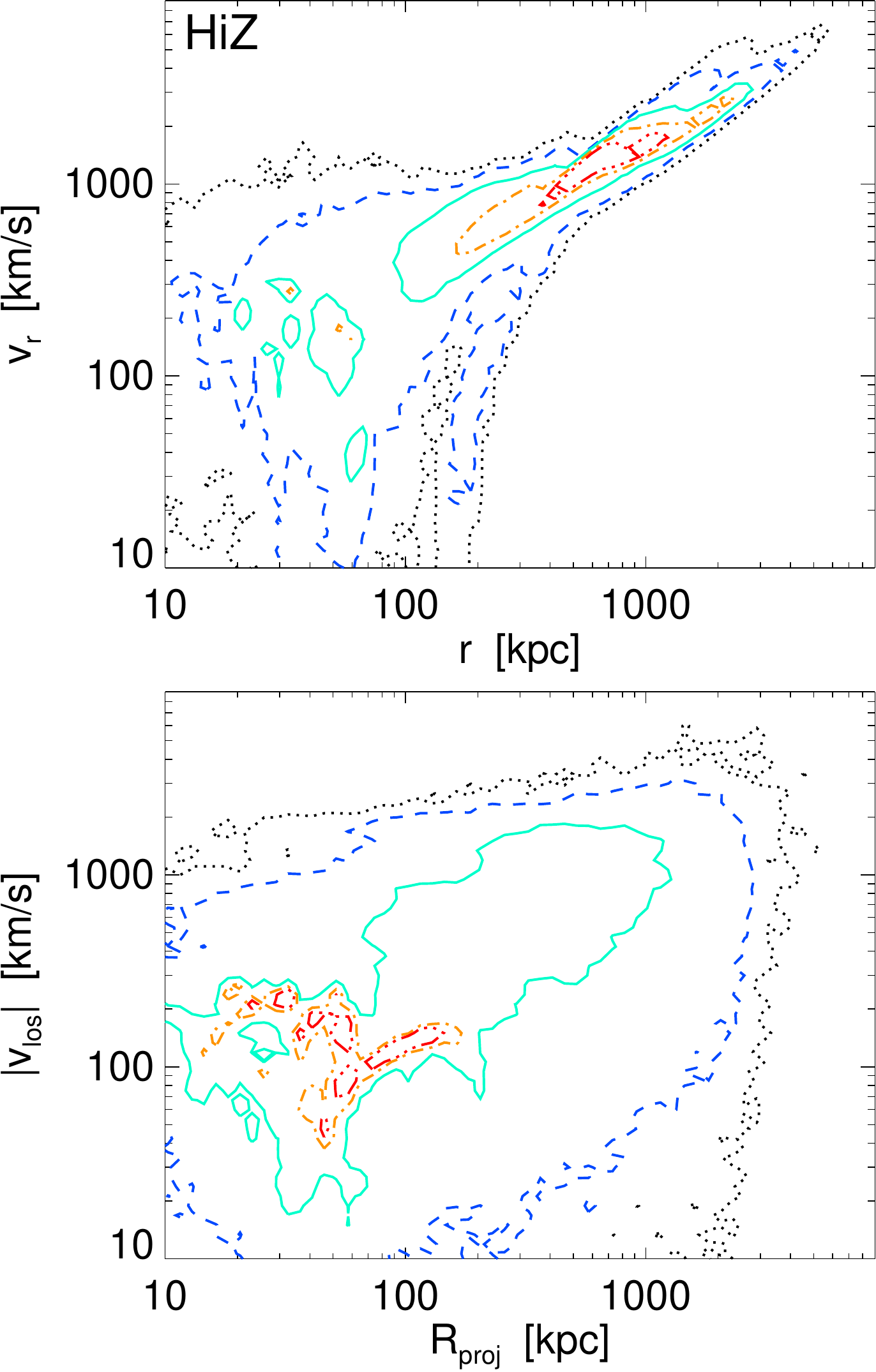}{1.0}
    \caption{{\em Top:} Distribution of radial outflow velocities ($v_{r}$) for one example simulation (the HiZ {\bf e} merger, chosen because it features the largest velocity range), as a function of three-dimensional distance ($r$) from the galaxy center, at the final time of the simulation. Contours are iso-density contours at $(0.001,\,0.01,\,0.1,\,0.3\,0.5)$ times maximum (in the plotted log-log coordinates). At smaller radii there is a broad range of outflow speed, along with inflow and rotationally supported material, but the median relation develops a Hubble-like flow. Fitting a power-law, we obtain $\langle v_{r} \rangle \propto r^{0.7-1.0}$. 
    {\em Bottom:} Same, but in projection. We plot the distribution of the absolute value of the line-of-sight velocity versus projected distance from the galaxy center (for a random viewing angle). The scatter is much larger owing to projection effects, but the scaling is similar $|v_{\rm los}| \propto R_{\rm proj}^{0.7-1.0}$.
    \label{fig:v.vs.r}}
\end{figure}

\vspace{-0.5cm}
\section{Velocity Structure of Outflows}
\label{sec:vel}

Figure~\ref{fig:wind.phases} shows that in all cases the winds have a broad velocity distribution extending to $>1000\,{\rm km\,s^{-1}}$, but most of the wind mass is near $\sim V_{c}$, with relatively little ($\ll1\%$ of the mass) at large $v\gtrsim 500\,{\rm km\,s^{-1}}$. Observationally, winds in bright ULIRGs have velocities similar to those here \citep{heckman:superwind.abs.kinematics,rupke:outflows,martin05:outflows.in.ulirgs}, but in AGN-dominated, late-stage mergers the wind velocities typically reach much larger values, $\sim 500-2000\,{\rm km\,s^{-1}}$ \citep[see e.g.][]{feruglio:2010.mrk231.agn.fb,sturm:2011.ulirg.herschel.outflows,rupke:2011.outflow.mrk231,greene:2011.nlr.outflows,greene:2012.quasar.outflow}.\footnote{Note that these are the outflows on {\em galactic} scales; there are much higher-velocity outflows still associated with broad absorption line systems and outflows near the AGN itself \citep[for a review, see][]{veilleux:winds}, although some of these may even have substantial components at $\sim$kpc scales \citep{dunn:agn.fb.from.strong.outflows,bautista:2010.strong.agn.fb}.} This suggests that wind velocity may be a useful observational discriminant between starburst and AGN driving mechanisms. We find that, without AGN feedback, the distribution of wind velocities below the escape velocity but above the disk circular velocity is quite flat; above this characteristic velocity however, it is exponentially decreasing with increasing $v_{r}$. 

Fig.~\ref{fig:vfield} shows the geometry of this velocity field on different scales, for a specific system (HiZ {\bf e}) at a given instant (pre-coalescence here). On scales comparable to the disks, it is clear that there are large non-radial components in the outflows, tracing both the orbital motions of the galaxies and the rotation of the disk (with local components from e.g.\ individual star clusters). Since the wind is driven from radii throughout the star-forming disk, this rotation component is detectable in the wind geometry out to $\sim100$\,kpc, consistent again with what has been seen in ULIRGs \citep{martin06:outflow.extend.origin,gauthier:2012.outflow.geometry}. On still larger scales, though, the outflow is primarily radial. 

Figs.~\ref{fig:integ.spec}, \ref{fig:ifu}, \&\ \ref{fig:v.vs.r} attempt to give an estimate of the observable line-of-sight velocity distribution. In Fig.~\ref{fig:integ.spec} we show the mass-weighted line-of-sight velocity distribution of all gas for each galaxy, at the snapshot nearest the peak in the starburst (generally shortly after nuclear coalescence), after projection onto a random axis. We stress that a direct model of e.g.\ an observed line profile requires properly modeling emission and absorption and involves three-dimensional line transport, which can differ significantly between lines and is outside of the scope of our study here \citep[but see e.g.][]{cooper:2009.wind.cloud.observability}. But this gives a rough guide to observed behaviors, especially if we consider different phases separately. In the cold gas especially, we see multiple narrow components with separations comparable to the circular velocity; these reflect unrelaxed merger kinematics. But in the cold/warm gas, and especially in the hot gas, we also see broad wings extending to a few hundred ${\rm km\,s^{-1}}$, a direct consequence of the winds. In Fig.~\ref{fig:ifu} we show the same, for just one system but spatially resolved across the halo. This makes obvious how the narrow component corresponds to different dense filaments and clumps$^{\ref{foot:clump.caveats}}$, but also shows how both the narrow offset and broad winds systematically vary across the galaxy (owing primarily to projection effects). The broad wings have a lower covering factor, but are more obvious out at large radii, as material has escaped to larger scales more quickly. Note the resemblance between the distributions here and those suggested in observed starbursts \citep[e.g.][]{martin:2009.outflow.accel.starburst,steidel:2010.outflow.kinematics,gauthier:2012.outflow.geometry}.

In Fig.~\ref{fig:v.vs.r}, we plot the radial outflow velocity of all gas versus three-dimensional distance from the galaxy center, at the end of one example simulation (i.e.\ at a post-starburst time, although the result is qualitatively similar at earlier times). The broad velocity distribution and mix of orientations/directions is obvious at smaller radii, as is the trend towards primarily radial outflow at the largest radii (since this is material which has escaped the galaxy). A Hubble-like flow develops quickly, and fitting a power-law to the median $v_{r}$ versus $r$ gives $v_{r}\propto r^{0.7-1.0}$. Some of the increase in velocity with distance owes to continuous acceleration, but most of the trend simply arises because the fastest-moving material escapes to the largest radii. We also show the velocity in projection, plotting the line-of-sight velocity versus projected distance for all gas. As expected, projection effects greatly broaden the distribution. However, the trend is similar -- fitting a power law $|v_{\rm los}|\propto R_{\rm proj}^{0.7-1.0}$ gives a similar scaling.

\begin{figure*}
    \centering
    \scaleup
    \plotsidesize{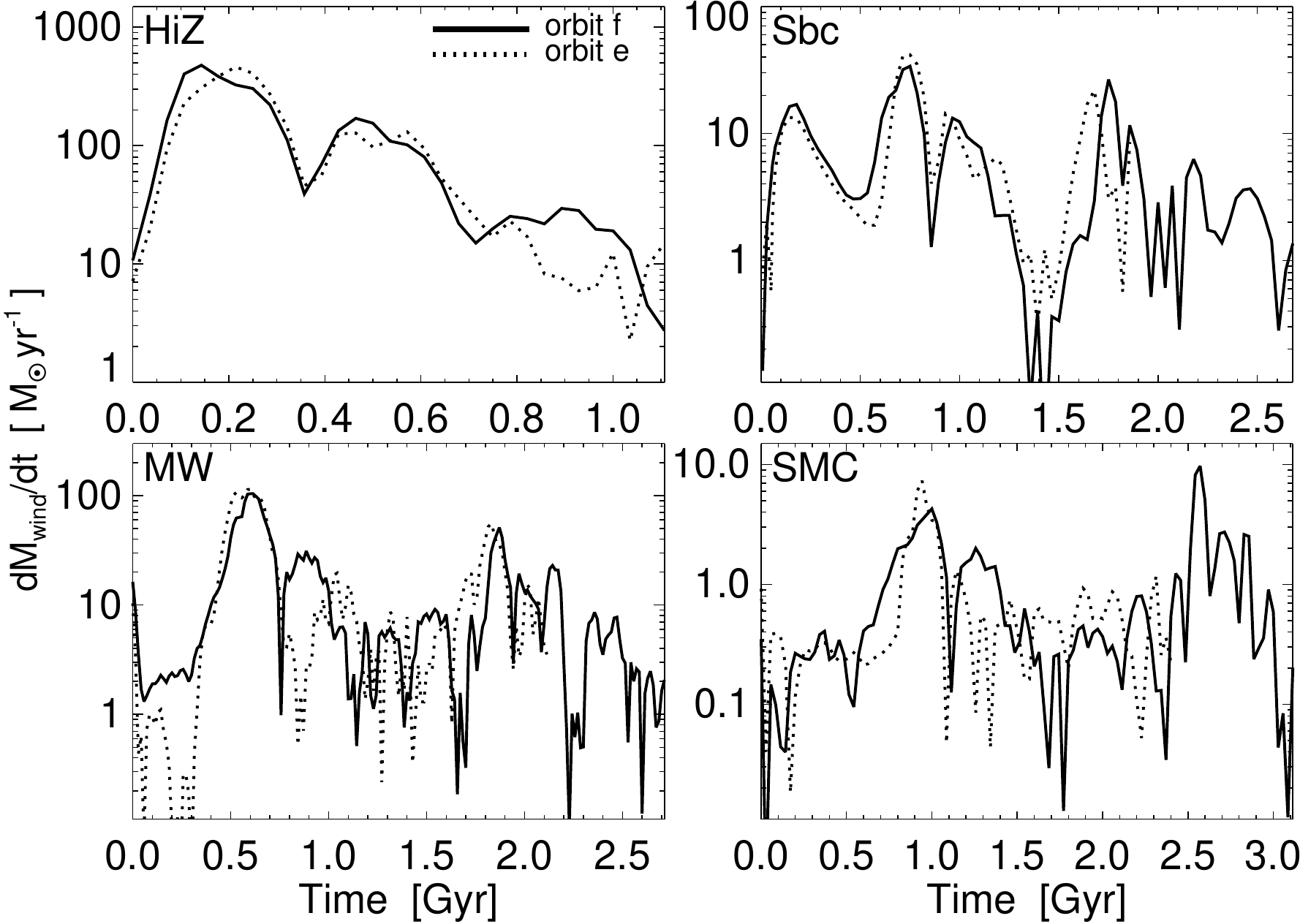}{0.8}
    \caption{Galactic super-wind mass outflow rates $\dot{M}_{\rm wind}$. We compare each disk model (labeled), and both orbits. The mass outflow rate is averaged over $\approx 2\times10^{7}\,$yr intervals, about a dynamical time; wind material is defined by a Bernoulli parameter $b>0$. The absolute outflow rates are highest in the HiZ case (up to $\sim600\,\msun\,{\rm yr^{-1}}$ over these timescales), but even in the SMC case reach $\sim10\,\msun\,{\rm yr^{-1}}$. The peaks broadly follow the starbursts at first couple passages and final coalescence. There is surprisingly little orbital dependence in the typical outflow rates. 
    \label{fig:winds.abs}}
\end{figure*}

\begin{figure*}
    \centering
    \scaleup
    \plotsidesize{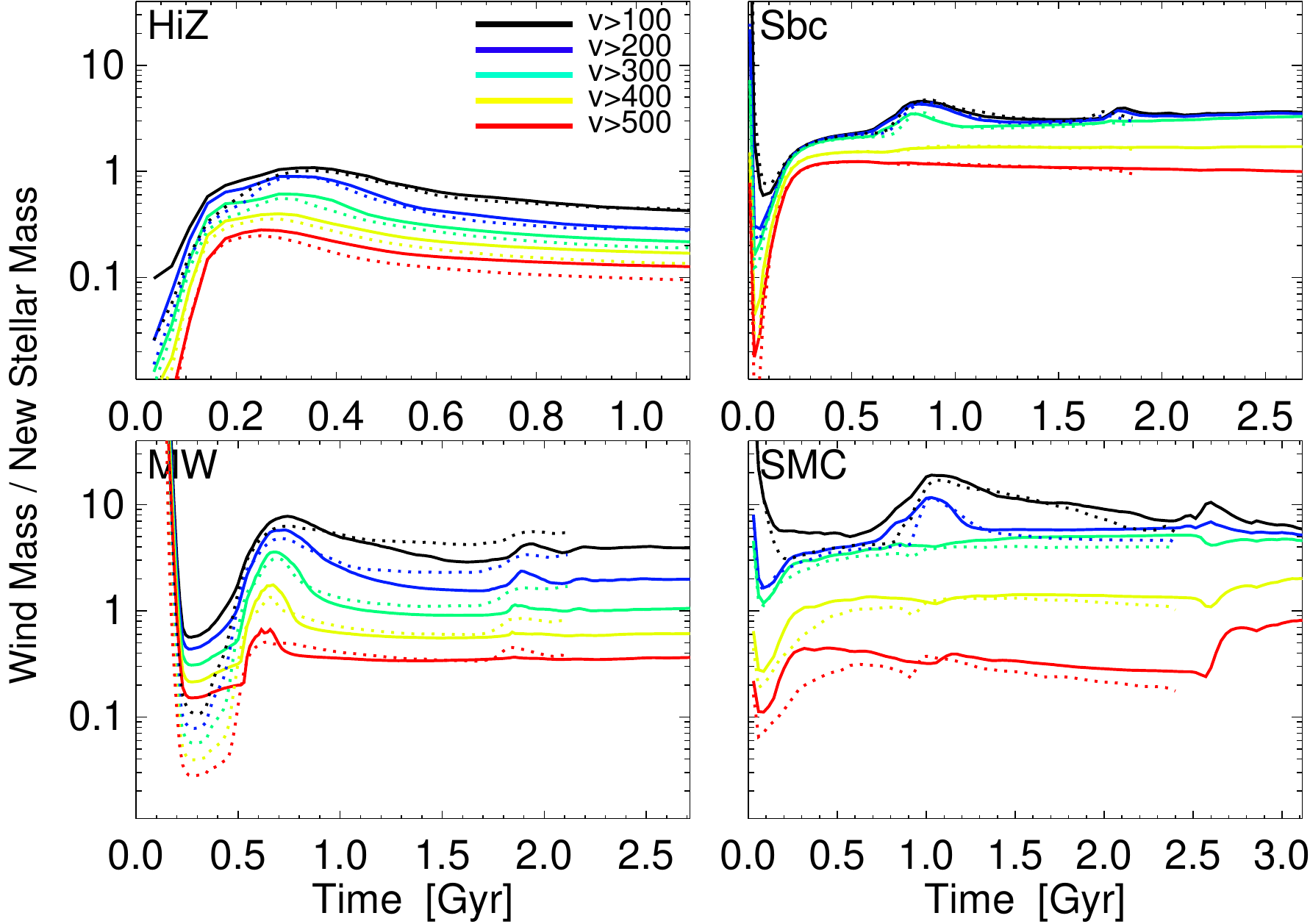}{0.8}
    \caption{Average galactic wind mass-loading efficiency ($\equiv M_{\rm wind}/M_{\rm new}$; 
    where $M_{\rm wind} = \int\,\dot{M}_{\rm wind}$ and $M_{\rm new} = \int\,\dot{M}_{\ast}$) 
    for each galaxy model ({\bf f}/{\bf e} orbits shown as solid/dotted lines as Fig.~\ref{fig:winds.abs}). 
    As shown in \paperthree, the mass-loading increases going from high-mass (HiZ, MW) systems to lower-mass (Sbc) and dwarf (SMC) galaxies. The mechanisms dominating the outflow also transition from radiation pressure to SNe, respectively. As isolated disks, 
    the HiZ, Sbc, MW, SMC models have $\dot{M}_{\rm wind}\sim(1,\,3-5,1-3,\,10-20)\,\dot{M}_{\ast}$. Similar values appear here, with weak time-dependence. Although the absolute outflow rates in starbursts in Fig.~\ref{fig:winds.abs} are very large, they follow from large SFRs; the mass loading efficiency is more sensitive to global galaxy properties than merger stage. We also consider the mass-loading of material with various cuts in the outflow radial velocity $v_{r}$ (in ${\rm km\,s^{-1}}$); the distribution of outflow velocities is very broad.$^{\ref{foot:wind.equilibrium}}$
    \label{fig:winds}}
\end{figure*}

\vspace{-0.5cm}
\section{Outflow Rates and Mass-Loading Efficiencies}
\label{sec:rates}

In Figures~\ref{fig:winds.abs} \&\ \ref{fig:winds}, we show the wind mass outflow rate during the mergers. Fig.~\ref{fig:winds.abs} shows it in absolute units. However the outflow rate is typically quantified in terms of its ratio to the star formation rate $\dot{M}_{\rm wind}/\dot{M}_{\ast}$, i.e.\ the wind mass per unit mass of stars formed, so we show this in Fig.~\ref{fig:winds}. In both cases, we define the ``wind'' as material with a positive Bernoulli parameter $b\equiv (v^{2} + 3\,c_{s}^{2} - v_{\rm esc}^{2})/2$, i.e., material that would escape in the absence of additional forces or cooling. In Fig.~\ref{fig:winds}, we show the wind mass above different absolute radial velocity cuts to highlight the characteristic velocities.\footnote{\label{foot:wind.equilibrium}Note that the first $\sim10^{8}$\,yr reflect both out-of-equilibrium initial conditions and contributions from the ``pre-existing'' stars in the initial conditions to the outflows (not the self-consistently formed stars in the simulation), so should be ignored).}

To first order, the wind mass-loading efficiency does not strongly depend on the merger stage. The {\em absolute} outflow rate in Fig.~\ref{fig:winds.abs} during the starburst is very large, $\sim1-10$ times the SFR for the models here, or $\sim10-500\,\msun\,{\rm yr^{-1}}$ in absolute units. But Fig.~\ref{fig:winds} shows that this is not proportionally much larger than what is seen for the isolated versions of these galaxy models;  $\dot{M}_{\rm wind}/\dot{M}_{\ast}$ is relatively flat in time. In some cases it even drops as the merger begins; this is because of the increase in surface densities making the escape of photons and hot gas needed to drive the winds less efficient. In \paperthree\ we parameterize the dependence of total wind mass-loading on galaxy properties (for isolated disks) as 
\be
\label{eqn:bestfit.eqn}
{\Bigl \langle}
\frac{\dot{M}_{\rm wind}}{\dot{M}_{\ast}}
{\Bigr \rangle}
\approx 
10\,
{\Bigl (} \frac{V_{\rm c}(R)}{100\,{\rm km\,s^{-1}}} {\Bigr)}^{-(1\pm0.3)}\,
{\Bigl (} \frac{\Sigma_{\rm gas}(R)}{10\,{\rm \msun\,{\rm pc}^{-2}}} {\Bigr)}^{-(0.5\pm0.15)}\,
\ee
with a scatter of $\sim50\%$ in normalization. 
This appears to provide a reasonable fit to the values in Fig.~\ref{fig:winds}. 

In the MW case, however, there does appear to be a sharp increase in the wind mass-loading at the first-passage and coalescence bursts. This is because, for that model, the gas fraction is small and so the specific SFR is much lower than any other model; the result shown in \paperthree\ is that the ``wind'' (in the isolated MW case) is mostly directly venting hot gas from SNe and stellar winds, rather than any entrained material ``blown out.'' It is, in short, below the threshold of SFR or ``feedback strength'' needed to blow out more material. But when the specific SFR is boosted in the merger passages, this allows it to drive efficient winds. This is not captured in the simple scaling for wind mass-loading efficiencies proposed for isolated galaxies in \paperthree.

\begin{figure}
    \centering
    \plotonesize{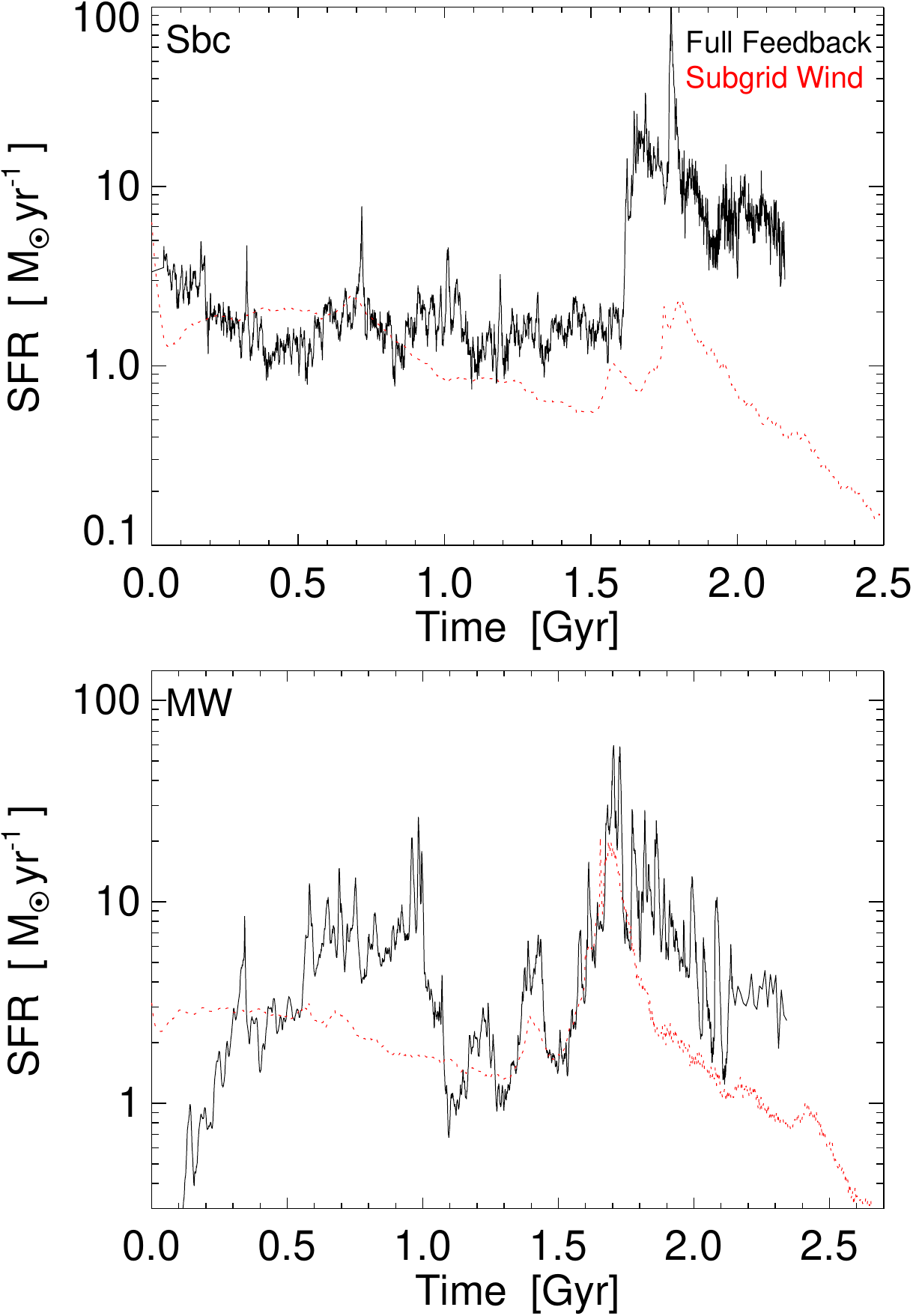}{1.0}
    \caption{Star formation rate versus time for our simulations Sbc and MW {\bf e} (``full feedback''; for the remaining simulations, see \paperfour\ Fig.~8). We compare to simulations run with identical initial conditions but a sub-grid model for both the ISM phase structure (``effective equation of state'') and winds (particles ``kicked out'' of the galaxy at fixed mass-loading, matched to the mean value in the corresponding ``full feedback'' simulation). The sub-grid wind model ``wipes out'' substructure, blowing out gas before it collects in the central kpc and dramatically suppresses the multi-phase starbursts. Also, because gas in the winds is completely ejected from the galaxy, the post-merger SFR decays too quickly in the sub-grid model. The difference is larger in the Sbc case because the sub-grid wind is assigned a larger mass-loading ($=5$, vs.\ $=1$ for the MW model), but evident in both.
    \label{fig:sfr}}
\end{figure}

\begin{figure}
    \centering
    \plotonesize{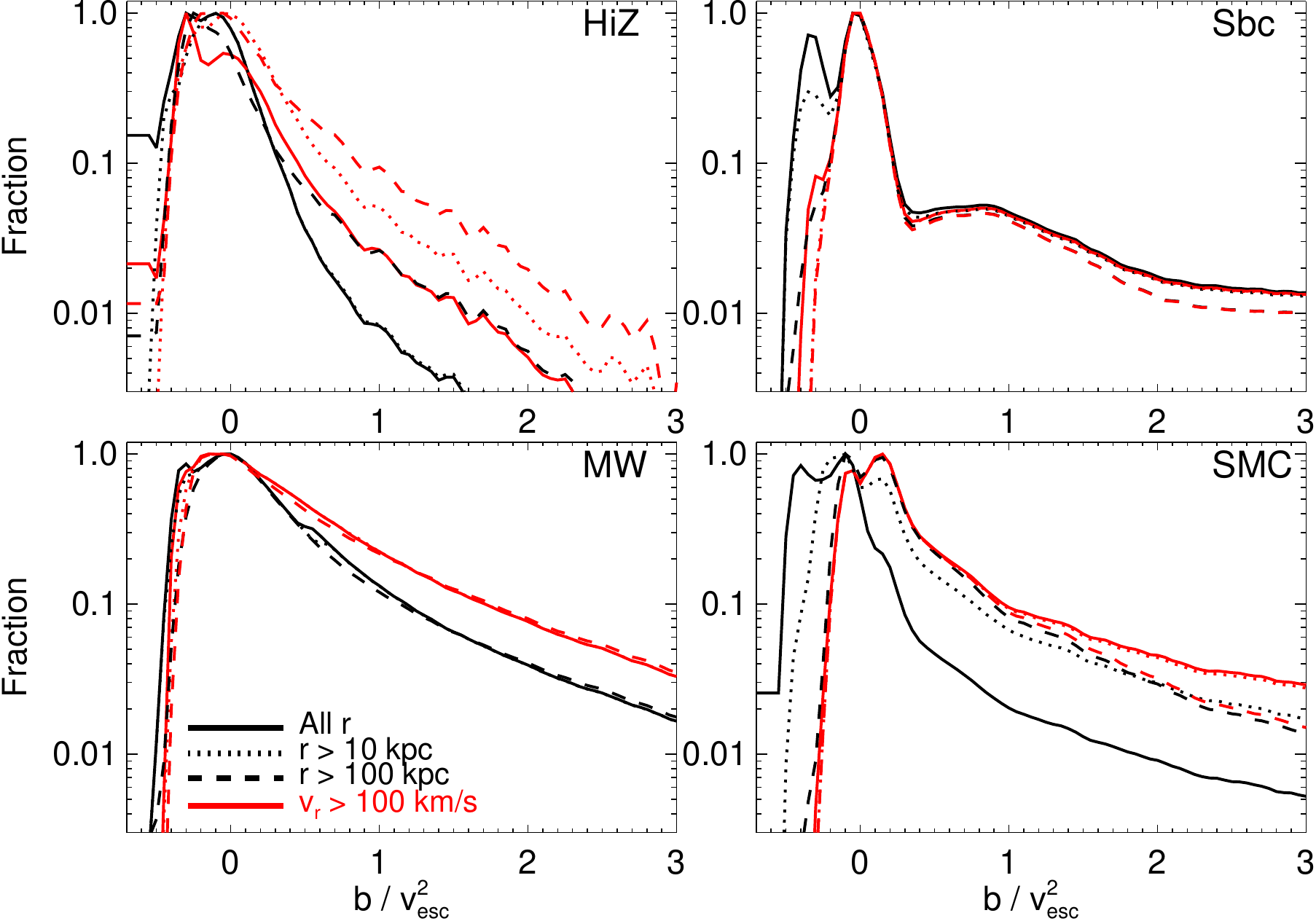}{1.}
    \caption{Distribution of Bernoulli parameters (binding energy) of the gas just after the peak of the merger-induced starburst. For each model ({\bf f} merger shown, but {\bf e} is very similar) we plot the distribution of mass per unit $b/v_{\rm esc}^{2}$, where $b\equiv (v^{2} + 3\,c_{s}^{2} - v_{\rm esc}^{2})/2$, so that $b>0$ corresponds to material which would escape in the absence of additional pressure forces and cooling. Black lines compare all gas, outside of different radii from the center of mass. Red lines correspond to the same limiting radii but restricted to gas with a radial outflow velocity $v_{r}>100\,{\rm km\,s^{-1}}$. There is clearly a large tail of $b>0$ un-bound gas in each case; the secondary peak at $b<0$ is virialized gas. But even for gas with large outflow speeds $>100\,{\rm km\,s^{-1}}$ and at radii $>10\,$kpc, there is a non-negligible ($\sim10-30\%$) fraction with $b<0$, which will fall back into the disk.
    \label{fig:wind.bernoulli.dist}}
\end{figure}

\vspace{-0.5cm}
\section{Effects of Outflows on Star Formation in Mergers}
\label{sec:sfr.fx}

The star formation properties of the simulations here are discussed in detail in \paperfour. However, here we wish to consider how these are affected by the outflows generated in the merger. In particular, in previous work on galaxy mergers (as discussed in \S~\ref{sec:intro}) it was not possible to explicitly resolve wind generation, and instead simulations used a variety of sub-grid approaches designed to model some ``effective'' wind scalings. We wish to examine how the consequences of these winds differ from those here.

Fig.~\ref{fig:sfr} shows the SFR versus time in a couple of our mergers, compared to simulations with identical initial conditions run using a simplified ``sub-grid'' treatment of stellar winds and the ISM from \citet{springel:multiphase}. In that model, rather than resolve the micro-structure of the ISM below $\sim$kpc scales, gas is assigned an ``effective equation of state'' (i.e.\ effective pressure above a low density threshold $=1\,{\rm cm^{-3}}$ where the medium is assumed to become multi-phase) motivated by the interplay of GMC formation and destruction, and turbulent driving and heating via stellar feedback. In the ``sub-grid'' treatment, the lack of resolution of star formation and feedback means that star formation is assigned statistically to gas at much lower densities, with an efficiency that must be tuned so that the model lies on the Kennicutt relation. Stellar winds are not resolved, so gas particles are instead stochastically kicked out of the galaxy at a rate proportional to the SFR, with a fixed velocity and mass-loading; they are then ``free-streamed'' (temporarily decoupled from the hydrodynamics) until they escape to a few kpc from the galaxy center (this simply guarantees that the assigned mass-loading is the actual wind mass, rather than most of it going into e.g.\ local turbulence). Here, we match the mass-loading to the mean measured in each corresponding simulation with the full treatment of feedback;\footnote{Though we caution that because of the differing implementations, it is difficult to exactly defined ``matched'' mass-loadings.} the velocity loading corresponds to a fraction $=0.1$ of SNe energy coupled, but is dynamically irrelevant because of the free-streaming condition. For simplicity, we compare just the Sbc {\bf e} and MW {\bf e} models, though the differences are robust across other simulations (the differences for the {\bf f} orbits are similar but less pronounced since the orbit produces a less extreme starburst).

We clearly see that the sub-grid wind models ``wipe out'' some of the starbursts at the time of coalescence. This occurs because the wind prescription blows material away at all densities, as the gas falls into the center; it thus effectively suppresses material actually getting into a dense, kpc-scale nucleus in the first place (rather than blowing out the material {\em after} the starburst begins). It also occurs because material cannot be ``saved'' for the starburst (this requires resolved phase structure), nor can dense portions of inflows ``resist'' the outflow occurring simultaneously. A consequence of this is that the mergers with certain simplified sub-grid wind models -- even of gas-rich disks on favorable orbits -- have difficulty reproducing ULIRGs and other extremely bright merging systems observed \citep[see e.g.][]{cox:feedback,dave:2010.smg.no.merger.cosmosims}. The effect appears more severe when gas fractions are large (the low-mass Sbc case), where the merger with sub-grid winds here barely produces an enhancement of the SFR..\footnote{Note that, in \paperfour, we compared the SFRs in the simulations here to those in models with an ``effective equation of state'' but {\em no} stellar wind model. In that case, the starbursts are more pronounced, since there is no wind to ``wipe out'' the inflow.}

In contrast, in the simulations with resolved feedback the cold, dense, compact gas being strongly torqued (and hence tightly bound) is difficult to entrain, and so more survives to contribute to the starburst, which then efficiently blows out less bound material recycled from inside clouds and/or being contributed from more diffuse flows at slightly later times. As shown in \paperfour\ and herein this simultaneously produces large SFR enhancements (comparable to ``sub-grid'' models with zero wind mass-loading) and ULIRG/Hyper-LIRG luminosities, while driving winds with large mass-loading factors. 

We also see that the post-merger ``tail'' of star formation is significantly different in sub-grid wind models. This is because the wind, by design, completely escapes the galaxy; meanwhile with no resolved feedback (merely effective pressure) in the disk, the material which remains is efficiently exhausted by star formation. In the simulations from \paperfour\ with no winds (but still using an effective equation-of-state model), nearly all the gas is efficiently exhausted (as expected). However, in the simulations with resolved feedback, there is a wide range of material in fountains with broad velocity, density, and temperature distributions (in addition to the unbound wind); in the best-studied systems, this appears to be observationally true as well \citep{strickland:2009.m82.wind.props}. Some of this then continuously rains down at later times onto the post-merger system. 

Fig.~\ref{fig:wind.bernoulli.dist} quantifies this by plotting the distribution of Bernoulli parameters $b$ for the gas, at the time just after peak starburst. There is a long tail towards $b\gg1$ which is the un-bound material, but even at large radial velocities $v_{r}>100\,{\rm km\,s^{-1}}$ and distances $>10\,$kpc from the disk, a non-trivial fraction of the wind material (tens of percent) is bound ($b<0$). This will rain down on the disk on a wide range of timescales corresponding to the broad velocity distribution, from $\sim100$\,Myr to several Gyr, and re-supplies the ``tail'' of star formation at late times. In contrast to simply concentrating the gas in one place at one time, this makes a complete ``shutoff'' of star formation much more difficult.

\vspace{-0.5cm}
\section{Discussion}
\label{sec:discussion}

In a series of papers, we have implemented detailed, explicit models for stellar feedback that can follow the formation and destruction of individual GMCs and star clusters: the models include star formation only at extremely high densities inside GMCs, and the mass, momentum, and energy flux from SNe (Types I \&\ II), stellar winds (both ``fast'' O-star winds and ``slow'' AGB winds), and radiation pressure (from UV/optical and multiply-scattered IR photons), as well as HII photoionization heating and molecular cooling. Here, we extend our models of isolated disks from previous papers to include major galaxy mergers. As a first study, we focus on simple, global properties, and compare them to those obtained from previous generations of simulations which did not follow these processes explicitly but instead adopted a simplified ``effective equation of state'' sub-grid model of the ISM.

With explicit feedback models, super-winds are generated in all passages with outflow rates from $\sim10-500\,\msun\,{\rm yr^{-1}}$. The simulated outflow rates are suggestively similar to observations of starburst winds in many observed ULIRGs and merging galaxies 
\citep[compare e.g.][]{heckman:1990.sb.superwinds,
heckman:superwind.abs.kinematics,
rupke:outflows,
coil:2011.postsb.winds,
grimes:fuse.starburst.obs.of.feedback,
rubin:2011.outflow.in.halo,
martin05:outflows.in.ulirgs,
martin06:outflow.extend.origin,
martin99:outflow.vs.m,
sato:2009.ulirg.outflows,
gauthier:2012.outflow.geometry}. 

Although the absolute outflow rates can be enormous in the starburst, we find that the mass-loading efficiency -- i.e.\ the outflow rate per unit star formation rate -- is broadly similar for each galaxy model to the isolated version of that system. In other words, outflow efficiencies -- on average -- appear to depend more strongly on global dynamical properties (escape velocity, effective radius) than on the instantaneous dynamical state of the system. The characteristic mass-loading efficiency scales inversely with the escape velocity of the systems, increasing from about unity in massive systems ($\sim10^{11}\,\msun$) to $\sim10-20$ in SMC-mass dwarfs. However since the absolute SFRs are much larger in massive systems, the absolute outflow rates tend to increase with galaxy mass (for gas-rich systems). We caution that, as for isolated systems in \paperthree, there is very large variability in the instantaneous mass-loading efficiency $\dot{M}_{\rm wind}/\dot{M}_{\ast}$ -- at least an order of magnitude scatter. 

We make predictions for the distribution of velocities, densities, and temperatures/phases of the outflows, which extend those shown in \paperthree\ for isolated disks. The distribution of velocities is broad, as shown for isolated systems in \paperthree, but extends characteristically to the escape velocity (a couple times the maximum circular velocity), with a long (but relatively low-mass) tail of material in higher-velocity components. This is also similar to observations -- typical velocities are a couple to a few hundred ${\rm km\,s^{-1}}$ (references above). The winds are characteristically multiphase, with a large fraction of the mass in each of the cold, warm ionized, and hot (pressure-supported) phase, and a broad range of densities spanning several orders of magnitude. The cold phases are predominantly accelerated by radiation pressure, and become more prominent in the more gas-rich, higher-density systems, including the central regions of the merger, while gas heated to high temperatures by SNe ejecta is more prominent in low-mass and/or gas-poor systems, and in the more diffuse (volume-filling) outflow. 

In contrast to isolated disks, the wind kinematics are (unsurprisingly) complex. The winds do not entirely originate in the nuclear starburst, but over the entire surface of the disks as the merger proceeds; not only do the extended disks contain non-trivial star formation (see \paperfour), but the material there (being lower-density, less tightly-bound, and less strongly-torqued) is easier to accelerate into the wind (or entrain in the outflow emerging from the very center). As such, near the disks the kinematics retain memory of the orbit and disk rotation (as well as more chaotic components from individual super-star clusters on sub-kpc scales), as has been mapped in detailed studies of local merging systems \citep{martin06:outflow.extend.origin}. On larger scales the outflow becomes primarily radial, but tends towards a unipolar or bipolar structure; however since the disk orientations are changing during the merger, this leads to multiple overlapping shells/bubbles with different orientations at different radii, also similar to observations \citep{gauthier:2012.outflow.geometry}. Because the winds contain material at a wide range of velocities, different outflows ``catch up'' to one another, and a complex three-dimensional structure develops. This also imprints a Hubble flow-like pattern of larger velocity material at larger radii (even without the effects, present here, of continuous thermal and radiation pressure acceleration), similar to that observed \citep[references above and][]{martin:2009.outflow.accel.starburst,steidel:2010.outflow.kinematics}. All gas phases include components extending to velocities of several hundred ${\rm km\,s^{-1}}$, but the cold/warm phases also show narrow features with smaller offsets $\sim100\,{\rm km\,s^{-1}}$ that indicate both merger kinematics and the structure of individual streams and entrained/accelerated shells of cool gas (though the breakup of these shells into some small cold clumps at large radii from the galaxy is subject to significant numerical caveats$^{\ref{foot:clump.caveats}}$), while the hot components show a smoother, broader velocity distribution. 

This multi-phase nature of the wind and its driving across the disk are critical to the fact that systems can {\em simultaneously} drive winds with large average mass-loading factors and also avoid ``wiping out'' all structure inside the disk -- including the kpc-scale gas concentrations that power the starburst itself! In contrast, sub-grid models which do not resolve the generation of winds but simply insert some mass-loading by hand can produce very different results. We show that some implementations of these subgrid models can suppress the merger-induced enhancement of the SFR -- which can (when mass-loadings are large) make it difficult to form ULIRGs or bright sub-millimeter galaxies. This may be related to some known difficulties reproducing these rare populations in cosmological simulations \citep[e.g.][]{dave:2010.smg.no.merger.cosmosims,hayward:2011.smg.merger.rt}. It is clear that careful treatment of subgrid wind models is necessary if one wishes to properly resolve the dynamics of gas flows and SFR enhancements within galaxies. 

It is also important that a significant fraction of the wind material is not unbound, and falls back into the disk over the couple Gyr after the final coalescence starburst, leading to a slow, extended ``tail'' in the starburst decay. Especially in prograde mergers, this can greatly enhance the magnitude and duration of post-merger star formation, relative to the older models which treat the ISM and feedback physics in a ``sub-grid'' manner. This has important implications for ``quenching'' of star formation in massive galaxies. If quenching were possible without the presence of some additional feedback source -- say, from an AGN -- then the simulations here are the most optimal case for this. They are cosmologically isolated galaxies, so there is zero new accretion; moreover, an equal-mass merger represents the most efficient means to exhaust a large amount of gas quickly via star formation, much moreso than an isolated disk \citep[see e.g.][]{hopkins:groups.qso,hopkins:groups.ell}. But we find that with the presence of stellar feedback, many of our models -- including the already gas-poor MW-like system, maintain post-merger SFRs nearly as large as their steady-state pre-merger SFR. The systems simulated here would take several Gyr to cross the ``green valley'' and turn red, much longer than the $<$Gyr quenching timescale required by observations \citep[see][]{martin:mass.flux,snyder:2011.ka.gal.sims}. Far from resolving this by gas expulsion, stellar feedback makes the ``quenching problem'' {\em harder}. As shown in \citet{moster:2011.gas.halo.merger.fx}, addition of realistic gas halos around the merging galaxies (even without continuous accretion) only further enhances the post-merger SFR. This is the short-timescale manifestation of a general problem in cosmological simulations; over a Hubble time, recycled material from galaxy progenitors is re-captured, and leads to large star formation rates and excessive stellar masses in high-mass galaxies \citep{oppenheimer:recycled.wind.accretion}. If gas is to be swept out of galaxies efficiently after a merger or starburst, the models imply that some other form of feedback -- perhaps from bright quasars -- is necessary. This is also suggested by observations of late-stage mergers, which find that in the AGN-dominated systems at quasar luminosities, outflow masses are enhanced and the outflow velocities reach $\sim1000\,{\rm km\,s^{-1}}$, larger than those we find driven by stellar feedback \citep{feruglio:2010.mrk231.agn.fb,tremonti:postsb.outflows,sturm:2011.ulirg.herschel.outflows,rupke:2011.outflow.mrk231}. These high velocities (well above the escape velocity) may provide a unique signature of AGN-driven outflows, since our simulations drive very little mass to such high values. But we caution that the relative timing of AGN and starbursts is uncertain (though the simulations suggest AGN follow the starburst; see \citealt{hopkins:sb.agn.delay}), and AGN (owing to their complicated duty cycles) may not still be active when AGN-driven winds are observable. 

In a companion paper, we will examine the star clusters formed in these simulations. The mass/luminosity distribution, spatial locations, formation time distribution, and physical properties of these clusters represent a powerful constraint on small-scale models of the ISM and star formation. 

We have also restricted our focus to major mergers. Studies of mergers with varying mass ratios suggest that the qualitative behaviors discussed here should not depend on mass ratio for ratios to about 3:1 or 4:1, and even at lower mass ratios they can be considered similar but with an ``efficiency'' of inducing starbursts and violent relaxation that scales approximately linearly with mass ratio \citep{hernquist.mihos:minor.mergers,cox:massratio.starbursts,naab:minor.mergers,younger:minor.mergers}. Since the simplest properties of the winds (e.g.\ their mass loading and characteristic velocities) seem to scale relatively simply between isolated disks and major mergers, we expect that minor mergers will represent an intermediate case. 

We note that recent studies comparing cosmological simulations done with {\small GADGET} and the new moving mesh code {\small AREPO} \citep{springel:arepo} have highlighted discrepancies between grid codes and smoothed particle hydrodynamics (SPH) in some problems related to galaxy formation in a cosmological context \citep{agertz:2007.sph.grid.mixing,vogelsberger:2011.arepo.vs.gadget.cosmo,sijacki:2011.gadget.arepo.hydro.tests,keres:2011.arepo.gadget.disk.angmom,bauer:2011.sph.vs.arepo.shocks,torrey:2011.arepo.disks}.  However, we have also performed idealized simulations of mergers between individual galaxies and found excellent agreement between {\small GADGET} and {\small AREPO} for e.g. gas-inflow rates, star formation histories, and the mass in the ensuing starbursts \citep{hayward:arepo.gadget.mergers}. Moreover, in \citet{hopkins:lagrangian.pressure.sph} we show that many of these discrepancies can be resolved with small modifications to the equations of motion, and test the differences in galaxies with the full feedback models presented here. Although subtle numerical issues can influence quantities like fluid mixing and hence hot gas cooling, we show that the SFRs agree very well and wind masses agree to within a factor of two, much smaller than the differences if we remove feedback. Finally, the differences in numerical methods are also minimized when the flows of interest are supersonic (as opposed to sub-sonic), which is very much the case here \citep{kitsionas:2009.grid.sph.compare.turbulence,price:2010.grid.sph.compare.turbulence,bauer:2011.sph.vs.arepo.shocks}. 

These new models allow us to follow the structure of the gas in the central regions of starburst systems at high resolution. This makes them an ideal laboratory to study feedback physics under extreme conditions in, say, the center of Arp 220 and other very dense galaxies. We have also, for clarity, neglected AGN feedback in these models, but we expect it may have a very significant effect on the systems after the final coalescence
\citep[e.g.][]{dimatteo:msigma,springel:red.galaxies,springel:models}. With high-resolution models that include the phase structure of the ISM, it becomes meaningful to include much more explicit physical models for AGN feedback.

\vspace{-0.1in}
\acknowledgments 
We thank Eliot Quataert for helpful discussions and contributions motivating this work, and thank the anonymous referee and editor for insightful comments and suggestions. 
Support for PFH was provided by NASA through Einstein Postdoctoral Fellowship Award Number PF1-120083 issued by the Chandra X-ray Observatory Center, which is operated by the Smithsonian Astrophysical Observatory for and on behalf of the NASA under contract NAS8-03060.
EQ is supported in part by NASA grant NNG06GI68G and 
the David and Lucile Packard Foundation.  
\\

\bibliography{/Users/phopkins/Documents/work/papers/ms}

\begin{thebibliography}{132}
\expandafter\ifx\csname natexlab\endcsname\relax\def\natexlab#1{#1}\fi

\bibitem[{{Agertz} {et~al.}(2007)}]{agertz:2007.sph.grid.mixing}
{Agertz}, O., {et~al.} 2007, \mnras, 380, 963

\bibitem[{{Aguirre} {et~al.}(2001){Aguirre}, {Hernquist}, {Schaye}, {Weinberg},
  {Katz}, \& {Gardner}}]{aguirre:2001.igm.metal.evol.sims}
{Aguirre}, A., {Hernquist}, L., {Schaye}, J., {Weinberg}, D.~H., {Katz}, N., \&
  {Gardner}, J. 2001, \apj, 560, 599

\bibitem[{{Bauer} \& {Springel}(2012)}]{bauer:2011.sph.vs.arepo.shocks}
{Bauer}, A., \& {Springel}, V. 2012, \mnras, 423, 3102

\bibitem[{{Bautista} {et~al.}(2010){Bautista}, {Dunn}, {Arav}, {Korista},
  {Moe}, \& {Benn}}]{bautista:2010.strong.agn.fb}
{Bautista}, M.~A., {Dunn}, J.~P., {Arav}, N., {Korista}, K.~T., {Moe}, M., \&
  {Benn}, C. 2010, \apj, 713, 25

\bibitem[{{Bournaud} {et~al.}(2011)}]{bournaud10}
{Bournaud}, F., {et~al.} 2011, \apj, 730, 4

\bibitem[{{Brook} {et~al.}(2011)}]{brook:2010.low.ang.mom.outflows}
{Brook}, C.~B., {et~al.} 2011, \mnras, 415, 1051

\bibitem[{Chapman {et~al.}(2009)Chapman, Blain, Ibata, Ivison, Smail, \&
  Morrison}]{chapman:submm.halo.clustering}
Chapman, S.~C., Blain, A., Ibata, R., Ivison, R.~J., Smail, I., \& Morrison, G.
  2009, The Astrophysical Journal, 691, 560

\bibitem[{{Chen} {et~al.}(2010){Chen}, {Tremonti}, {Heckman}, {Kauffmann},
  {Weiner}, {Brinchmann}, \& {Wang}}]{chen:2010.local.outflow.properties}
{Chen}, Y.-M., {Tremonti}, C.~A., {Heckman}, T.~M., {Kauffmann}, G., {Weiner},
  B.~J., {Brinchmann}, J., \& {Wang}, J. 2010, \aj, 140, 445

\bibitem[{{Coil} {et~al.}(2011){Coil}, {Weiner}, {Holz}, {Cooper}, {Yan}, \&
  {Aird}}]{coil:2011.postsb.winds}
{Coil}, A.~L., {Weiner}, B.~J., {Holz}, D.~E., {Cooper}, M.~C., {Yan}, R., \&
  {Aird}, J. 2011, \apj, 743, 46

\bibitem[{Cole {et~al.}(2000)Cole, Lacey, Baugh, \&
  Frenk}]{cole:durham.sam.initial}
Cole, S., Lacey, C.~G., Baugh, C.~M., \& Frenk, C.~S. 2000, \mnras, 319, 168

\bibitem[{{Cooper} {et~al.}(2009){Cooper}, {Bicknell}, {Sutherland}, \&
  {Bland-Hawthorn}}]{cooper:2009.wind.cloud.observability}
{Cooper}, J.~L., {Bicknell}, G.~V., {Sutherland}, R.~S., \& {Bland-Hawthorn},
  J. 2009, \apj, 703, 330

\bibitem[{{Cox} {et~al.}(2006{\natexlab{a}}){Cox}, {Dutta}, {Di Matteo},
  {Hernquist}, {Hopkins}, {Robertson}, \& {Springel}}]{cox:kinematics}
{Cox}, T.~J., {Dutta}, S.~N., {Di Matteo}, T., {Hernquist}, L., {Hopkins},
  P.~F., {Robertson}, B., \& {Springel}, V. 2006{\natexlab{a}}, \apj, 650, 791

\bibitem[{{Cox} {et~al.}(2006{\natexlab{b}}){Cox}, {Jonsson}, {Primack}, \&
  {Somerville}}]{cox:feedback}
{Cox}, T.~J., {Jonsson}, P., {Primack}, J.~R., \& {Somerville}, R.~S.
  2006{\natexlab{b}}, \mnras, 373, 1013

\bibitem[{{Cox} {et~al.}(2008){Cox}, {Jonsson}, {Somerville}, {Primack}, \&
  {Dekel}}]{cox:massratio.starbursts}
{Cox}, T.~J., {Jonsson}, P., {Somerville}, R.~S., {Primack}, J.~R., \& {Dekel},
  A. 2008, \mnras, 384, 386

\bibitem[{{Cullen} \& {Dehnen}(2010)}]{cullen:2010.inviscid.sph}
{Cullen}, L., \& {Dehnen}, W. 2010, \mnras, 408, 669

\bibitem[{{Dasyra} {et~al.}(2006)}]{dasyra:mass.ratio.conditions}
{Dasyra}, K.~M., {et~al.} 2006, \apj, 638, 745

\bibitem[{{Dasyra} {et~al.}(2007)}]{dasyra:pg.qso.dynamics}
---. 2007, \apj, 657, 102

\bibitem[{{Dav{\'e}} {et~al.}(2010){Dav{\'e}}, {Finlator}, {Oppenheimer},
  {Fardal}, {Katz}, {Kere{\v s}}, \&
  {Weinberg}}]{dave:2010.smg.no.merger.cosmosims}
{Dav{\'e}}, R., {Finlator}, K., {Oppenheimer}, B.~D., {Fardal}, M., {Katz}, N.,
  {Kere{\v s}}, D., \& {Weinberg}, D.~H. 2010, \mnras, 404, 1355

\bibitem[{{Dehnen} \& {Aly}(2012)}]{dehnen.aly:2012.sph.kernels}
{Dehnen}, W., \& {Aly}, H. 2012, \mnras, 425, 1068

\bibitem[{{Di Matteo} {et~al.}(2005){Di Matteo}, {Springel}, \&
  {Hernquist}}]{dimatteo:msigma}
{Di Matteo}, T., {Springel}, V., \& {Hernquist}, L. 2005, \nat, 433, 604

\bibitem[{{Doyon} {et~al.}(1994){Doyon}, {Wells}, {Wright}, {Joseph}, {Nadeau},
  \& {James}}]{Doyon94}
{Doyon}, R., {Wells}, M., {Wright}, G.~S., {Joseph}, R.~D., {Nadeau}, D., \&
  {James}, P.~A. 1994, \apjl, 437, L23

\bibitem[{{Dunn} {et~al.}(2010)}]{dunn:agn.fb.from.strong.outflows}
{Dunn}, J.~P., {et~al.} 2010, \apj, 709, 611

\bibitem[{{Durier} \& {Dalla Vecchia}(2012)}]{durier:2012.timestep.limiter}
{Durier}, F., \& {Dalla Vecchia}, C. 2012, \mnras, 419, 465

\bibitem[{{Evans}(1999)}]{evans:1999.sf.gmc.review}
{Evans}, II, N.~J. 1999, \araa, 37, 311

\bibitem[{{Faucher-Gigu{\`e}re} {et~al.}(2008){Faucher-Gigu{\`e}re}, {Lidz},
  {Hernquist}, \& {Zaldarriaga}}]{faucher:ion.background.evol}
{Faucher-Gigu{\`e}re}, C.-A., {Lidz}, A., {Hernquist}, L., \& {Zaldarriaga}, M.
  2008, \apj, 688, 85

\bibitem[{{Feruglio} {et~al.}(2010){Feruglio}, {Maiolino}, {Piconcelli},
  {Menci}, {Aussel}, {Lamastra}, \& {Fiore}}]{feruglio:2010.mrk231.agn.fb}
{Feruglio}, C., {Maiolino}, R., {Piconcelli}, E., {Menci}, N., {Aussel}, H.,
  {Lamastra}, A., \& {Fiore}, F. 2010, \aap, 518, L155

\bibitem[{{Gao} \& {Solomon}(2004)}]{gao:2004.hcn.compilation}
{Gao}, Y., \& {Solomon}, P.~M. 2004, \apjs, 152, 63

\bibitem[{{Gauthier} \& {Chen}(2012)}]{gauthier:2012.outflow.geometry}
{Gauthier}, J.-R., \& {Chen}, H.-W. 2012, \mnras, 424, 1952

\bibitem[{{Genzel} {et~al.}(2001){Genzel}, {Tacconi}, {Rigopoulou}, {Lutz}, \&
  {Tecza}}]{Genzel01}
{Genzel}, R., {Tacconi}, L.~J., {Rigopoulou}, D., {Lutz}, D., \& {Tecza}, M.
  2001, \apj, 563, 527

\bibitem[{{Governato} {et~al.}(2010)}]{governato:2010.dwarf.gal.form}
{Governato}, F., {et~al.} 2010, \nat, 463, 203

\bibitem[{{Greene} {et~al.}(2011){Greene}, {Zakamska}, {Ho}, \&
  {Barth}}]{greene:2011.nlr.outflows}
{Greene}, J.~E., {Zakamska}, N.~L., {Ho}, L.~C., \& {Barth}, A.~J. 2011, \apj,
  732, 9

\bibitem[{{Greene} {et~al.}(2012){Greene}, {Zakamska}, \&
  {Smith}}]{greene:2012.quasar.outflow}
{Greene}, J.~E., {Zakamska}, N.~L., \& {Smith}, P.~S. 2012, \apj, 746, 86

\bibitem[{Grimes {et~al.}(2009)}]{grimes:fuse.starburst.obs.of.feedback}
Grimes, J.~P., {et~al.} 2009, The Astrophysical Journal Supplement, 181, 272

\bibitem[{{Guo} {et~al.}(2010){Guo}, {White}, {Li}, \&
  {Boylan-Kolchin}}]{guo:2010.hod.constraints}
{Guo}, Q., {White}, S., {Li}, C., \& {Boylan-Kolchin}, M. 2010, \mnras, 404,
  1111

\bibitem[{{Hayward} {et~al.}(2011{\natexlab{a}}){Hayward}, {Kere{\v s}},
  {Jonsson}, {Narayanan}, {Cox}, \& {Hernquist}}]{hayward:2011.smg.merger.rt}
{Hayward}, C.~C., {Kere{\v s}}, D., {Jonsson}, P., {Narayanan}, D., {Cox},
  T.~J., \& {Hernquist}, L. 2011{\natexlab{a}}, \apj, 743, 159

\bibitem[{{Hayward}
  {et~al.}(2011{\natexlab{b}})}]{hayward:arepo.gadget.mergers}
{Hayward}, C.~C., {et~al.} 2011{\natexlab{b}}, \mnras, in preparation

\bibitem[{{Heckman} {et~al.}(1990){Heckman}, {Armus}, \&
  {Miley}}]{heckman:1990.sb.superwinds}
{Heckman}, T.~M., {Armus}, L., \& {Miley}, G.~K. 1990, \apjs, 74, 833

\bibitem[{{Heckman} {et~al.}(2000){Heckman}, {Lehnert}, {Strickland}, \&
  {Armus}}]{heckman:superwind.abs.kinematics}
{Heckman}, T.~M., {Lehnert}, M.~D., {Strickland}, D.~K., \& {Armus}, L. 2000,
  \apjs, 129, 493

\bibitem[{{Hernquist}(1990)}]{hernquist:profile}
{Hernquist}, L. 1990, \apj, 356, 359

\bibitem[{{Hernquist} \& {Mihos}(1995)}]{hernquist.mihos:minor.mergers}
{Hernquist}, L., \& {Mihos}, J.~C. 1995, \apj, 448, 41

\bibitem[{{Hernquist} {et~al.}(1993){Hernquist}, {Spergel}, \&
  {Heyl}}]{hernquist:phasespace}
{Hernquist}, L., {Spergel}, D.~N., \& {Heyl}, J.~S. 1993, \apj, 416, 415

\bibitem[{{Hopkins}(2011)}]{hopkins:sb.agn.delay}
{Hopkins}, P.~F. 2011, \mnras, L376

\bibitem[{{Hopkins}(2012{\natexlab{a}})}]{hopkins:excursion.ism}
---. 2012{\natexlab{a}}, \mnras, 423, 2016

\bibitem[{{Hopkins}(2012{\natexlab{b}})}]{hopkins:excursion.imf}
---. 2012{\natexlab{b}}, \mnras, 423, 2037

\bibitem[{{Hopkins}(2013)}]{hopkins:lagrangian.pressure.sph}
---. 2013, \mnras, 428, 2840

\bibitem[{{Hopkins} {et~al.}(2009{\natexlab{a}}){Hopkins}, {Cox}, {Dutta},
  {Hernquist}, {Kormendy}, \& {Lauer}}]{hopkins:cusps.ell}
{Hopkins}, P.~F., {Cox}, T.~J., {Dutta}, S.~N., {Hernquist}, L., {Kormendy},
  J., \& {Lauer}, T.~R. 2009{\natexlab{a}}, \apjs, 181, 135

\bibitem[{{Hopkins} {et~al.}(2013){Hopkins}, {Cox}, {Hernquist}, {Narayanan},
  {Hayward}, \& {Murray}}]{hopkins:stellar.fb.mergers}
{Hopkins}, P.~F., {Cox}, T.~J., {Hernquist}, L., {Narayanan}, D., {Hayward},
  C.~C., \& {Murray}, N. 2013, \mnras, 430, 1901

\bibitem[{{Hopkins} {et~al.}(2008{\natexlab{a}}){Hopkins}, {Cox}, {Kere{\v s}},
  \& {Hernquist}}]{hopkins:groups.ell}
{Hopkins}, P.~F., {Cox}, T.~J., {Kere{\v s}}, D., \& {Hernquist}, L.
  2008{\natexlab{a}}, \apjs, 175, 390

\bibitem[{{Hopkins} \& {Hernquist}(2010)}]{hopkins:sb.ir.lfs}
{Hopkins}, P.~F., \& {Hernquist}, L. 2010, \mnras, 402, 985

\bibitem[{{Hopkins} {et~al.}(2008{\natexlab{b}}){Hopkins}, {Hernquist}, {Cox},
  {Dutta}, \& {Rothberg}}]{hopkins:cusps.mergers}
{Hopkins}, P.~F., {Hernquist}, L., {Cox}, T.~J., {Dutta}, S.~N., \& {Rothberg},
  B. 2008{\natexlab{b}}, \apj, 679, 156

\bibitem[{{Hopkins} {et~al.}(2008{\natexlab{c}}){Hopkins}, {Hernquist}, {Cox},
  \& {Kere{\v s}}}]{hopkins:groups.qso}
{Hopkins}, P.~F., {Hernquist}, L., {Cox}, T.~J., \& {Kere{\v s}}, D.
  2008{\natexlab{c}}, \apjs, 175, 356

\bibitem[{{Hopkins} {et~al.}(2009{\natexlab{b}}){Hopkins}, {Lauer}, {Cox},
  {Hernquist}, \& {Kormendy}}]{hopkins:cores}
{Hopkins}, P.~F., {Lauer}, T.~R., {Cox}, T.~J., {Hernquist}, L., \& {Kormendy},
  J. 2009{\natexlab{b}}, \apjs, 181, 486

\bibitem[{{Hopkins} {et~al.}(2012{\natexlab{a}}){Hopkins}, {Narayanan},
  {Quataert}, \& {Murray}}]{hopkins:dense.gas.tracers}
{Hopkins}, P.~F., {Narayanan}, D., {Quataert}, E., \& {Murray}, N.
  2012{\natexlab{a}}, \mnras, in press, arXiv:1209.0459

\bibitem[{{Hopkins} {et~al.}(2011{\natexlab{a}}){Hopkins}, {Quataert}, \&
  {Murray}}]{hopkins:binding.sf.prescription}
{Hopkins}, P.~F., {Quataert}, E., \& {Murray}, N. 2011{\natexlab{a}}, \mnras,
  in prep

\bibitem[{{Hopkins} {et~al.}(2011{\natexlab{b}}){Hopkins}, {Quataert}, \&
  {Murray}}]{hopkins:rad.pressure.sf.fb}
---. 2011{\natexlab{b}}, \mnras, 417, 950

\bibitem[{{Hopkins} {et~al.}(2012{\natexlab{b}}){Hopkins}, {Quataert}, \&
  {Murray}}]{hopkins:stellar.fb.winds}
---. 2012{\natexlab{b}}, \mnras, 421, 3522

\bibitem[{{Hopkins} {et~al.}(2012{\natexlab{c}}){Hopkins}, {Quataert}, \&
  {Murray}}]{hopkins:fb.ism.prop}
---. 2012{\natexlab{c}}, \mnras, 421, 3488

\bibitem[{{James} {et~al.}(1999){James}, {Bate}, {Wells}, {Wright}, \&
  {Doyon}}]{James99}
{James}, P., {Bate}, C., {Wells}, M., {Wright}, G., \& {Doyon}, R. 1999,
  \mnras, 309, 585

\bibitem[{{Joseph} \& {Wright}(1985)}]{joseph85}
{Joseph}, R.~D., \& {Wright}, G.~S. 1985, \mnras, 214, 87

\bibitem[{{Juneau} {et~al.}(2009){Juneau}, {Narayanan}, {Moustakas}, {Shirley},
  {Bussmann}, {Kennicutt}, \& {Vanden
  Bout}}]{juneau:2009.enhanced.dense.gas.ulirgs}
{Juneau}, S., {Narayanan}, D.~T., {Moustakas}, J., {Shirley}, Y.~L.,
  {Bussmann}, R.~S., {Kennicutt}, Jr., R.~C., \& {Vanden Bout}, P.~A. 2009,
  \apj, 707, 1217

\bibitem[{{Katz} {et~al.}(1996){Katz}, {Weinberg}, \&
  {Hernquist}}]{katz:treesph}
{Katz}, N., {Weinberg}, D.~H., \& {Hernquist}, L. 1996, \apjs, 105, 19

\bibitem[{{Kere{\v s}} {et~al.}(2009){Kere{\v s}}, {Katz}, {Dav{\'e}},
  {Fardal}, \& {Weinberg}}]{keres:fb.constraints.from.cosmo.sims}
{Kere{\v s}}, D., {Katz}, N., {Dav{\'e}}, R., {Fardal}, M., \& {Weinberg},
  D.~H. 2009, \mnras, 396, 2332

\bibitem[{{Kere{\v s}} {et~al.}(2012){Kere{\v s}}, {Vogelsberger}, {Sijacki},
  {Springel}, \& {Hernquist}}]{keres:2011.arepo.gadget.disk.angmom}
{Kere{\v s}}, D., {Vogelsberger}, M., {Sijacki}, D., {Springel}, V., \&
  {Hernquist}, L. 2012, \mnras, 425, 2027

\bibitem[{{Kitsionas}
  {et~al.}(2009)}]{kitsionas:2009.grid.sph.compare.turbulence}
{Kitsionas}, S., {et~al.} 2009, \aap, 508, 541

\bibitem[{{Kroupa}(2002)}]{kroupa:imf}
{Kroupa}, P. 2002, Science, 295, 82

\bibitem[{{Krumholz} \& {Gnedin}(2011)}]{krumholz:2011.molecular.prescription}
{Krumholz}, M.~R., \& {Gnedin}, N.~Y. 2011, \apj, 729, 36

\bibitem[{{Lake} \& {Dressler}(1986)}]{LakeDressler86}
{Lake}, G., \& {Dressler}, A. 1986, \apj, 310, 605

\bibitem[{{Macci{\`o}} {et~al.}(2012){Macci{\`o}}, {Stinson}, {Brook},
  {Wadsley}, {Couchman}, {Shen}, {Gibson}, \&
  {Quinn}}]{maccio:2012.cuspcore.outflows}
{Macci{\`o}}, A.~V., {Stinson}, G., {Brook}, C.~B., {Wadsley}, J., {Couchman},
  H.~M.~P., {Shen}, S., {Gibson}, B.~K., \& {Quinn}, T. 2012, \apjl, 744, L9

\bibitem[{{Mannucci} {et~al.}(2006){Mannucci}, {Della Valle}, \&
  {Panagia}}]{mannucci:2006.snIa.rates}
{Mannucci}, F., {Della Valle}, M., \& {Panagia}, N. 2006, \mnras, 370, 773

\bibitem[{{Martin}(1999)}]{martin99:outflow.vs.m}
{Martin}, C.~L. 1999, \apj, 513, 156

\bibitem[{{Martin}(2005)}]{martin05:outflows.in.ulirgs}
---. 2005, \apj, 621, 227

\bibitem[{{Martin}(2006)}]{martin06:outflow.extend.origin}
---. 2006, \apj, 647, 222

\bibitem[{{Martin} \& {Bouch{\'e}}(2009)}]{martin:2009.outflow.accel.starburst}
{Martin}, C.~L., \& {Bouch{\'e}}, N. 2009, \apj, 703, 1394

\bibitem[{{Martin} {et~al.}(2010){Martin}, {Scannapieco}, {Ellison}, {Hennawi},
  {Djorgovski}, \& {Fournier}}]{martin:2010.metal.enriched.regions}
{Martin}, C.~L., {Scannapieco}, E., {Ellison}, S.~L., {Hennawi}, J.~F.,
  {Djorgovski}, S.~G., \& {Fournier}, A.~P. 2010, \apj, 721, 174

\bibitem[{{Martin} {et~al.}(2007)}]{martin:mass.flux}
{Martin}, D.~C., {et~al.} 2007, \apjs, 173, 342

\bibitem[{{Moster} {et~al.}(2011){Moster}, {Macci{\`o}}, {Somerville}, {Naab},
  \& {Cox}}]{moster:2011.gas.halo.merger.fx}
{Moster}, B.~P., {Macci{\`o}}, A.~V., {Somerville}, R.~S., {Naab}, T., \&
  {Cox}, T.~J. 2011, \mnras, 415, 3750

\bibitem[{{Naab} \& {Burkert}(2003)}]{naab:minor.mergers}
{Naab}, T., \& {Burkert}, A. 2003, \apj, 597, 893

\bibitem[{{Nagamine}(2010)}]{nagamine:2010.dwarf.gal.cosmo.review}
{Nagamine}, K. 2010, Advances in Astronomy, 2010

\bibitem[{{Narayanan} {et~al.}(2008){Narayanan}, {Cox}, \&
  {Hernquist}}]{narayanan:2008.sfr.densegas.corr}
{Narayanan}, D., {Cox}, T.~J., \& {Hernquist}, L. 2008, \apjl, 681, L77

\bibitem[{{Narayanan} {et~al.}(2006){Narayanan}, {Cox}, {Robertson},
  {Dav{\'e}}, {Di Matteo}, {Hernquist}, {Hopkins}, {Kulesa}, \&
  {Walker}}]{narayanan:co.outflows}
{Narayanan}, D., {Cox}, T.~J., {Robertson}, B., {Dav{\'e}}, R., {Di Matteo},
  T., {Hernquist}, L., {Hopkins}, P., {Kulesa}, C., \& {Walker}, C.~K. 2006,
  \apjl, 642, L107

\bibitem[{{Nelson} {et~al.}(2013){Nelson}, {Vogelsberger}, {Genel}, {Sijacki},
  {Kere{\v s}}, {Springel}, \&
  {Hernquist}}]{nelson:2013.arepo.coldflow.structure}
{Nelson}, D., {Vogelsberger}, M., {Genel}, S., {Sijacki}, D., {Kere{\v s}}, D.,
  {Springel}, V., \& {Hernquist}, L. 2013, \mnras, 429, 3353

\bibitem[{{Newman} {et~al.}(2012)}]{newman:z2.clump.winds}
{Newman}, S.~F., {et~al.} 2012, \apj, 752, 111

\bibitem[{{Oppenheimer} \& {Dav{\'e}}(2006)}]{oppenheimer:outflow.enrichment}
{Oppenheimer}, B.~D., \& {Dav{\'e}}, R. 2006, \mnras, 373, 1265

\bibitem[{{Oppenheimer} {et~al.}(2010){Oppenheimer}, {Dav{\'e}}, {Kere{\v s}},
  {Fardal}, {Katz}, {Kollmeier}, \&
  {Weinberg}}]{oppenheimer:recycled.wind.accretion}
{Oppenheimer}, B.~D., {Dav{\'e}}, R., {Kere{\v s}}, D., {Fardal}, M., {Katz},
  N., {Kollmeier}, J.~A., \& {Weinberg}, D.~H. 2010, \mnras, 406, 2325

\bibitem[{{Pakmor} \&
  {Springel}(2012)}]{pakmor:2012.mag.field.disk.evol.weak.fx}
{Pakmor}, R., \& {Springel}, V. 2012, \mnras, in press, arXiv:1212.1452

\bibitem[{{Papovich} {et~al.}(2005){Papovich}, {Dickinson}, {Giavalisco},
  {Conselice}, \& {Ferguson}}]{papovich:highz.sb.gal.timescales}
{Papovich}, C., {Dickinson}, M., {Giavalisco}, M., {Conselice}, C.~J., \&
  {Ferguson}, H.~C. 2005, \apj, 631, 101

\bibitem[{{Pettini} {et~al.}(2003){Pettini}, {Madau}, {Bolte}, {Prochaska},
  {Ellison}, \& {Fan}}]{pettini:2003.igm.metal.evol}
{Pettini}, M., {Madau}, P., {Bolte}, M., {Prochaska}, J.~X., {Ellison}, S.~L.,
  \& {Fan}, X. 2003, \apj, 594, 695

\bibitem[{{Powell} {et~al.}(2011){Powell}, {Slyz}, \&
  {Devriendt}}]{powell:2010.sne.fb.weak.winds}
{Powell}, L.~C., {Slyz}, A., \& {Devriendt}, J. 2011, \mnras, 414, 3671

\bibitem[{{Price} \&
  {Federrath}(2010)}]{price:2010.grid.sph.compare.turbulence}
{Price}, D.~J., \& {Federrath}, C. 2010, \mnras, 406, 1659

\bibitem[{{Read} \& {Hayfield}(2012)}]{read:2012.sph.w.dissipation.switches}
{Read}, J.~I., \& {Hayfield}, T. 2012, \mnras, 422, 3037

\bibitem[{{Robertson} {et~al.}(2006){Robertson}, {Cox}, {Hernquist}, {Franx},
  {Hopkins}, {Martini}, \& {Springel}}]{robertson:fp}
{Robertson}, B., {Cox}, T.~J., {Hernquist}, L., {Franx}, M., {Hopkins}, P.~F.,
  {Martini}, P., \& {Springel}, V. 2006, \apj, 641, 21

\bibitem[{{Rothberg} \& {Joseph}(2004)}]{rj:profiles}
{Rothberg}, B., \& {Joseph}, R.~D. 2004, \aj, 128, 2098

\bibitem[{{Rothberg} \& {Joseph}(2006)}]{rothberg.joseph:kinematics}
---. 2006, \aj, 131, 185

\bibitem[{{Rubin} {et~al.}(2011){Rubin}, {Prochaska}, {M{\'e}nard}, {Murray},
  {Kasen}, {Koo}, \& {Phillips}}]{rubin:2011.outflow.in.halo}
{Rubin}, K.~H.~R., {Prochaska}, J.~X., {M{\'e}nard}, B., {Murray}, N., {Kasen},
  D., {Koo}, D.~C., \& {Phillips}, A.~C. 2011, \apj, 728, 55

\bibitem[{{Rupke} {et~al.}(2005){Rupke}, {Veilleux}, \&
  {Sanders}}]{rupke:outflows}
{Rupke}, D.~S., {Veilleux}, S., \& {Sanders}, D.~B. 2005, \apj, 632, 751

\bibitem[{{Rupke} \& {Veilleux}(2011)}]{rupke:2011.outflow.mrk231}
{Rupke}, D.~S.~N., \& {Veilleux}, S. 2011, \apjl, 729, L27+

\bibitem[{{Saitoh} \& {Makino}(2009)}]{saitoh.makino:2009.timestep.limiter}
{Saitoh}, T.~R., \& {Makino}, J. 2009, \apjl, 697, L99

\bibitem[{{Sanders} \& {Mirabel}(1996)}]{sanders96:ulirgs.mergers}
{Sanders}, D.~B., \& {Mirabel}, I.~F. 1996, \araa, 34, 749

\bibitem[{{Sargent} {et~al.}(1987){Sargent}, {Sanders}, {Scoville}, \&
  {Soifer}}]{sargent87}
{Sargent}, A.~I., {Sanders}, D.~B., {Scoville}, N.~Z., \& {Soifer}, B.~T. 1987,
  \apjl, 312, L35

\bibitem[{{Sato} {et~al.}(2009){Sato}, {Martin}, {Noeske}, {Koo}, \&
  {Lotz}}]{sato:2009.ulirg.outflows}
{Sato}, T., {Martin}, C.~L., {Noeske}, K.~G., {Koo}, D.~C., \& {Lotz}, J.~M.
  2009, \apj, 696, 214

\bibitem[{{Schinnerer}
  {et~al.}(2008)}]{schinnerer:submm.merger.w.compact.mol.gas}
{Schinnerer}, E., {et~al.} 2008, \apjl, 689, L5

\bibitem[{{Scoville} {et~al.}(1986){Scoville}, {Sanders}, {Sargent}, {Soifer},
  {Scott}, \& {Lo}}]{scoville86}
{Scoville}, N.~Z., {Sanders}, D.~B., {Sargent}, A.~I., {Soifer}, B.~T.,
  {Scott}, S.~L., \& {Lo}, K.~Y. 1986, \apjl, 311, L47

\bibitem[{{Shier} \& {Fischer}(1998)}]{ShierFischer98}
{Shier}, L.~M., \& {Fischer}, J. 1998, \apj, 497, 163

\bibitem[{{Sijacki} {et~al.}(2012){Sijacki}, {Vogelsberger}, {Keres},
  {Springel}, \& {Hernquist}}]{sijacki:2011.gadget.arepo.hydro.tests}
{Sijacki}, D., {Vogelsberger}, M., {Keres}, D., {Springel}, V., \& {Hernquist},
  L. 2012, \mnras, 424, 2999

\bibitem[{{Snyder} {et~al.}(2011){Snyder}, {Cox}, {Hayward}, {Hernquist}, \&
  {Jonsson}}]{snyder:2011.ka.gal.sims}
{Snyder}, G.~F., {Cox}, T.~J., {Hayward}, C.~C., {Hernquist}, L., \& {Jonsson},
  P. 2011, \apj, 741, 77

\bibitem[{{Somerville} \& {Primack}(1999)}]{somerville99:sam}
{Somerville}, R.~S., \& {Primack}, J.~R. 1999, \mnras, 310, 1087

\bibitem[{{Songaila}(2005)}]{songaila:2005.igm.metal.evol}
{Songaila}, A. 2005, \aj, 130, 1996

\bibitem[{{Springel}(2005)}]{springel:gadget}
{Springel}, V. 2005, \mnras, 364, 1105

\bibitem[{Springel(2010)}]{springel:arepo}
Springel, V. 2010, \mnras, 401, 791

\bibitem[{{Springel} {et~al.}(2005{\natexlab{a}}){Springel}, {Di Matteo}, \&
  {Hernquist}}]{springel:red.galaxies}
{Springel}, V., {Di Matteo}, T., \& {Hernquist}, L. 2005{\natexlab{a}}, \apjl,
  620, L79

\bibitem[{{Springel} {et~al.}(2005{\natexlab{b}}){Springel}, {Di Matteo}, \&
  {Hernquist}}]{springel:models}
---. 2005{\natexlab{b}}, \mnras, 361, 776

\bibitem[{{Springel} \& {Hernquist}(2002)}]{springel:entropy}
{Springel}, V., \& {Hernquist}, L. 2002, \mnras, 333, 649

\bibitem[{{Springel} \& {Hernquist}(2003{\natexlab{a}})}]{springel:multiphase}
---. 2003{\natexlab{a}}, \mnras, 339, 289

\bibitem[{{Springel} \& {Hernquist}(2003{\natexlab{b}})}]{springel:lcdm.sfh}
---. 2003{\natexlab{b}}, \mnras, 339, 312

\bibitem[{{Steidel} {et~al.}(2010){Steidel}, {Erb}, {Shapley}, {Pettini},
  {Reddy}, {Bogosavljevi{\'c}}, {Rudie}, \&
  {Rakic}}]{steidel:2010.outflow.kinematics}
{Steidel}, C.~C., {Erb}, D.~K., {Shapley}, A.~E., {Pettini}, M., {Reddy}, N.,
  {Bogosavljevi{\'c}}, M., {Rudie}, G.~C., \& {Rakic}, O. 2010, \apj, 717, 289

\bibitem[{{Strickland} \& {Heckman}(2009)}]{strickland:2009.m82.wind.props}
{Strickland}, D.~K., \& {Heckman}, T.~M. 2009, \apj, 697, 2030

\bibitem[{{Sturm} {et~al.}(2011)}]{sturm:2011.ulirg.herschel.outflows}
{Sturm}, E., {et~al.} 2011, \apjl, 733, L16+

\bibitem[{{Tacconi} {et~al.}(2002){Tacconi}, {Genzel}, {Lutz}, {Rigopoulou},
  {Baker}, {Iserlohe}, \& {Tecza}}]{tacconi:ulirgs.sb.profiles}
{Tacconi}, L.~J., {Genzel}, R., {Lutz}, D., {Rigopoulou}, D., {Baker}, A.~J.,
  {Iserlohe}, C., \& {Tecza}, M. 2002, \apj, 580, 73

\bibitem[{{Tacconi} {et~al.}(2006)}]{tacconi:smg.maximal.sb.sizes}
{Tacconi}, L.~J., {et~al.} 2006, \apj, 640, 228

\bibitem[{{Tacconi} {et~al.}(2008)}]{tacconi:smg.mgr.lifetime.to.quiescent}
---. 2008, \apj, 680, 246

\bibitem[{{Teyssier} {et~al.}(2013){Teyssier}, {Pontzen}, {Dubois}, \&
  {Read}}]{teyssier:2013.cuspcore.outflow}
{Teyssier}, R., {Pontzen}, A., {Dubois}, Y., \& {Read}, J.~I. 2013, \mnras,
  429, 3068

\bibitem[{{Torrey} {et~al.}(2012){Torrey}, {Vogelsberger}, {Sijacki},
  {Springel}, \& {Hernquist}}]{torrey:2011.arepo.disks}
{Torrey}, P., {Vogelsberger}, M., {Sijacki}, D., {Springel}, V., \&
  {Hernquist}, L. 2012, \mnras, 427, 2224

\bibitem[{{Tremonti} {et~al.}(2007){Tremonti}, {Moustakas}, \&
  {Diamond-Stanic}}]{tremonti:postsb.outflows}
{Tremonti}, C.~A., {Moustakas}, J., \& {Diamond-Stanic}, A.~M. 2007, \apjl,
  663, L77

\bibitem[{{Uhlig} {et~al.}(2012){Uhlig}, {Pfrommer}, {Sharma}, {Nath},
  {En{\ss}lin}, \& {Springel}}]{uhlig:2012.cosmic.ray.streaming.winds}
{Uhlig}, M., {Pfrommer}, C., {Sharma}, M., {Nath}, B.~B., {En{\ss}lin}, T.~A.,
  \& {Springel}, V. 2012, \mnras, 423, 2374

\bibitem[{{Veilleux} {et~al.}(2005){Veilleux}, {Cecil}, \&
  {Bland-Hawthorn}}]{veilleux:winds}
{Veilleux}, S., {Cecil}, G., \& {Bland-Hawthorn}, J. 2005, \araa, 43, 769

\bibitem[{{Vogelsberger} {et~al.}(2012){Vogelsberger}, {Sijacki}, {Keres},
  {Springel}, \& {Hernquist}}]{vogelsberger:2011.arepo.vs.gadget.cosmo}
{Vogelsberger}, M., {Sijacki}, D., {Keres}, D., {Springel}, V., \& {Hernquist},
  L. 2012, \mnras, 425, 3024

\bibitem[{{White} \& {Frenk}(1991)}]{white:1991.galform}
{White}, S.~D.~M., \& {Frenk}, C.~S. 1991, \apj, 379, 52

\bibitem[{{Wiersma} {et~al.}(2009){Wiersma}, {Schaye}, \&
  {Smith}}]{wiersma:2009.coolingtables}
{Wiersma}, R.~P.~C., {Schaye}, J., \& {Smith}, B.~D. 2009, \mnras, 393, 99

\bibitem[{{Williams} \& {McKee}(1997)}]{williams:1997.gmc.prop}
{Williams}, J.~P., \& {McKee}, C.~F. 1997, \apj, 476, 166

\bibitem[{{Wolfire} {et~al.}(1995){Wolfire}, {Hollenbach}, {McKee}, {Tielens},
  \& {Bakes}}]{wolfire:1995.neutral.ism.phases}
{Wolfire}, M.~G., {Hollenbach}, D., {McKee}, C.~F., {Tielens}, A.~G.~G.~M., \&
  {Bakes}, E.~L.~O. 1995, \apj, 443, 152

\bibitem[{{Younger} {et~al.}(2008{\natexlab{a}}){Younger}, {Hopkins}, {Cox}, \&
  {Hernquist}}]{younger:minor.mergers}
{Younger}, J.~D., {Hopkins}, P.~F., {Cox}, T.~J., \& {Hernquist}, L.
  2008{\natexlab{a}}, \apj, 686, 815

\bibitem[{{Younger} {et~al.}(2008{\natexlab{b}})}]{younger:smg.sizes}
{Younger}, J.~D., {et~al.} 2008{\natexlab{b}}, \apj, 688, 59

\end{thebibliography}

\clearpage

\begin{appendix}
\vspace{-0.5cm}
\section{Numerical Tests of the SPH Method and Wind Phase Structure}
\label{sec:appendix:altsph}

\begin{figure}
    \plotonesize{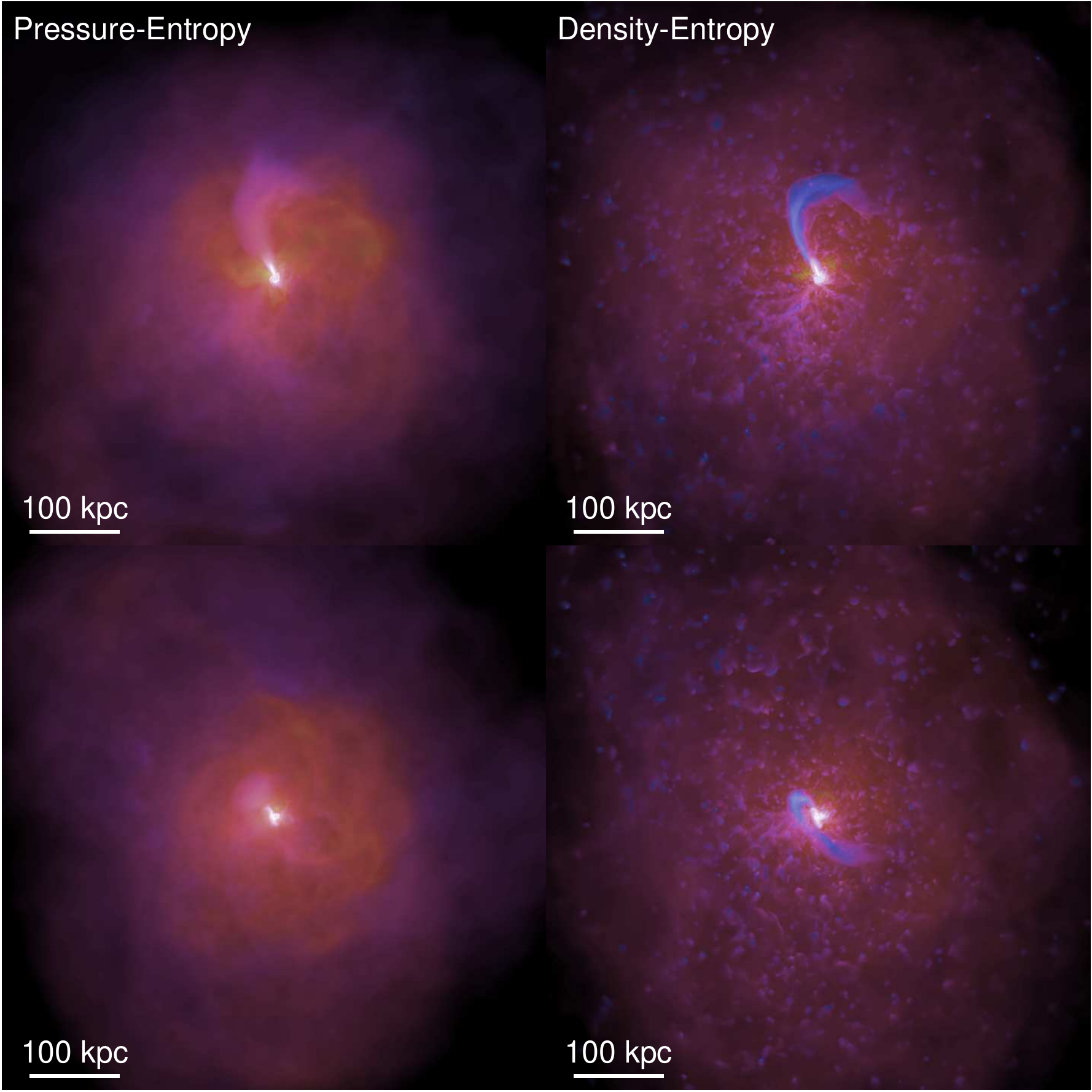}{1}
    \caption{Morphology of the gas in the merging SMC {\bf f} model, with colors encoding temperature as Fig.~\ref{fig:wind.images}. We compare two projections of the same simulation (at an identical time near coalescence). The ``density-entropy'' case is the standard formulation of the SPH equations of motion, our ``default'' model in the paper. The ``pressure-entropy'' case has been re-run with a new SPH code using the alternative ``pressure-entropy'' formulation of SPH developed in \citet{hopkins:lagrangian.pressure.sph}, which resolves some known problems in the ``standard'' SPH method treating fluid mixing instabilities and contact discontinuities (e.g.\ the Kelvin-Helmholtz instability). The new code also features an improved SPH smoothing kernel, artificial viscosity scheme, and includes UV background heating and self-shielding. The morphology of the hot gas shells/bubbles, warm outflowing filaments/shells at $\sim10-100\,$kpc and tidal tails is very similar. However, the previously cool gas at large radii ($\gg10\,$kpc) is now photo-heated to ``warm'' temperatures (in e.g.\ the tidal tails). The biggest difference is that the small, cold clumps at the largest radii (which were never self-gravitating and appeared to form in the outflow in the density-entropy models) do not appear in the new code. They are a numerical artifact of the ``density-entropy'' equation of motion. 
    \label{fig:smc.mgr.morph}}
\end{figure}

\begin{figure*}
    \plotsidesize{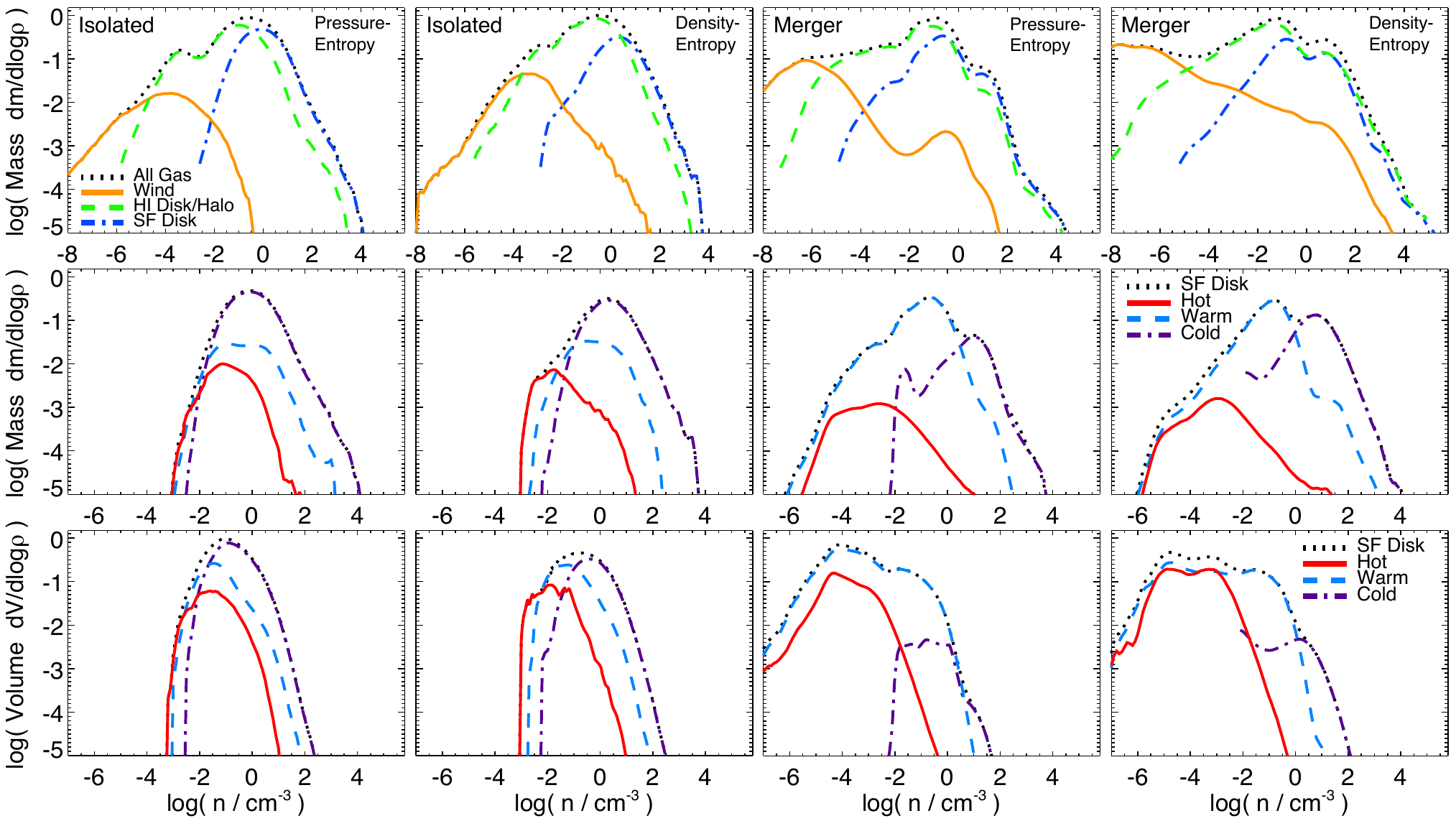}{1}
    \caption{Phase structure and density distribution of the gas in different components in the SMC {\bf f} merger simulations in Fig.~\ref{fig:smc.mgr.morph} and additional simulations of the isolated (non-merging) progenitor SMC disks of the merger simulation. The distributions are in the style of Fig.~\ref{fig:rho.dist}. {\em Top:} Mass-weighted density PDF for all gas in the simulation, and the gas within the multi-phase star-forming disk, the extended ionized disks and halo, and wind/outflow material. {\em Middle:} Mass-weighted density PDF within the star forming disk, divided into the cold/warm/hot phases. {\em Bottom:} Volume-weighted density PDF in the star-forming disk. In each case, both the merger and isolated disk are run with our ``default'' density-entropy formulation from the text and re-run with the improved pressure-entropy formulation from \citet{hopkins:lagrangian.pressure.sph}. As suggested from the Figures, the pressure-entropy formulation leads to more mixing that reduces the ``hot'' wind mass by a factor $\sim2$. The phase distribution {\em within the star forming disk} is nearly identical (down to $\lesssim1\%$ within the tails of the distributions) -- we stress that the elimination of the cold ``blobs'' at large radii in the outflow in Fig.~\ref{fig:smc.mgr.morph} does not apply to the GMCs and star clusters forming within the disk (which are well-resolved and self-gravitating, unlike the cold blobs in outflow). Note that the merger run with the pressure-entropy formulation self-consistently includes UV background heating, so the cold gas distribution self-consistently ``cuts off'' at a density $\lesssim 0.01\,{\rm cm^{-3}}$, very similar to where our post-processing estimates truncate the distribution in the text. Differences in the artificial viscosity and other numerical parameters appear to have small effects here. 
    \label{fig:rho.dist.vs.sph}}
\end{figure*}

\begin{figure*}
    \plotsidesize{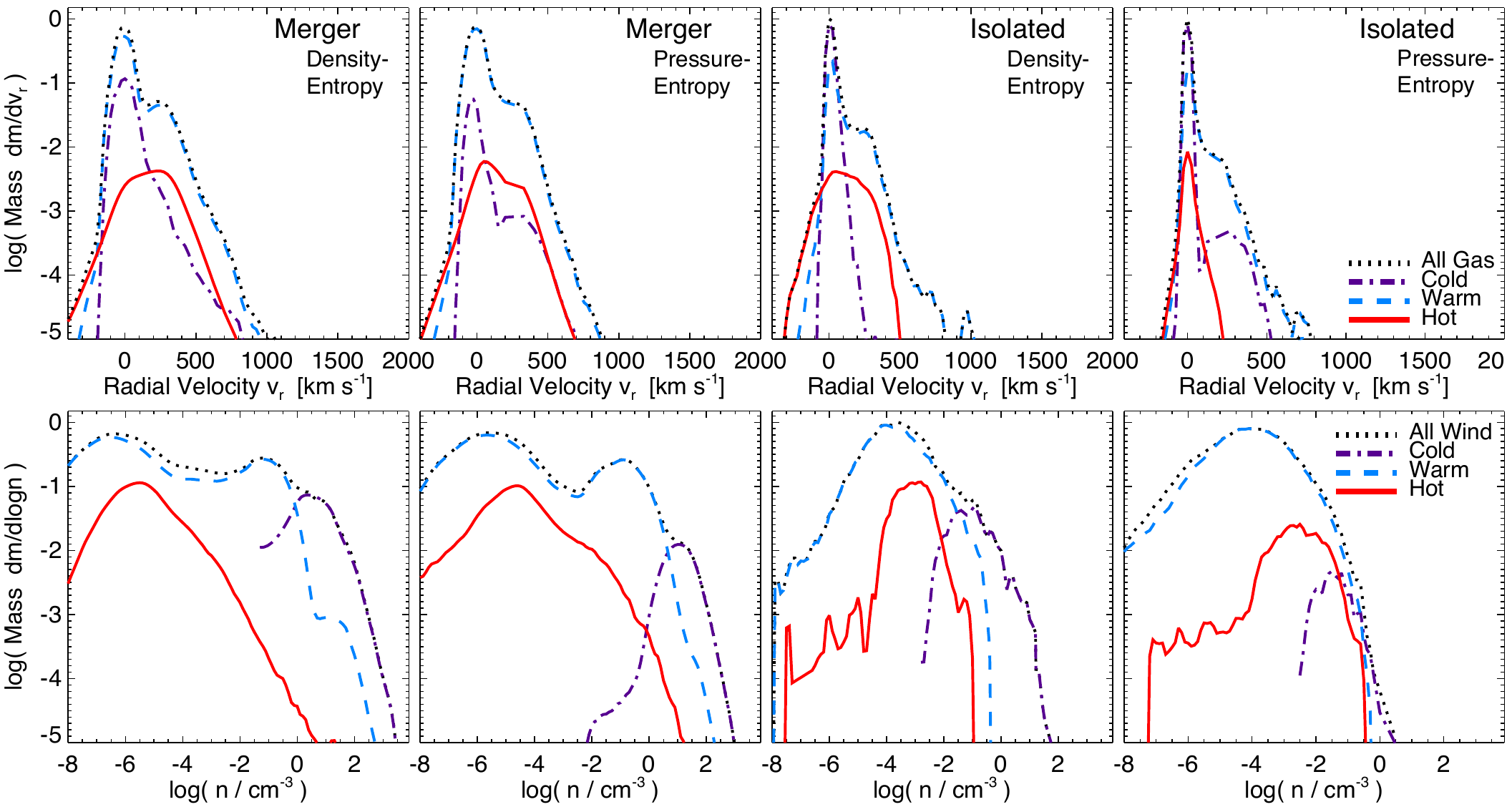}{1}
    \caption{Phase structure, velocity, and density distributions of the winds of the same simulations as Fig.~\ref{fig:rho.dist.vs.sph}, in the style of Fig.~\ref{fig:wind.phases}. {\em Top:} Mass-weighted distribution of gas outflow radial velocities, for all gas and divided into cold/warm/hot phases. 
    {\em Bottom:} Mass-weighted density distribution of material in the wind, divided by phases. 
    Again we compare the isolated SMC model and SMC {\bf f} merger, at the same instant in time, but with either our default ``density-entropy'' SPH or the revised ``pressure-entropy'' SPH. The resulting differences in the winds are larger than in the star-forming disks in Fig.~\ref{fig:rho.dist.vs.sph}. However they are still qualitatively similar. The temperature and density distributions have similar peaks and widths, but differ in their tails. As expected from the morphologies in Fig.~\ref{fig:smc.mgr.morph}, the cold ``blobs'' at large radii are more efficiently mixed with the warm outflow in the pressure-entropy formulation, leading to a smaller cold gas contribution in the wind (but this is not dominant, in any case). In the merger simulation, this mixing also enhances the cooling of the hot gas, so it is reduced as well (making the warm gas component even more dominant). In the merger, the total outflow mass at this stage is systematically lower in the pressure-entropy formulation by about $\sim40\%$. 
    \label{fig:wind.phases.vs.sph}}
\end{figure*}

\begin{figure}
    \centering
    \plotonesize{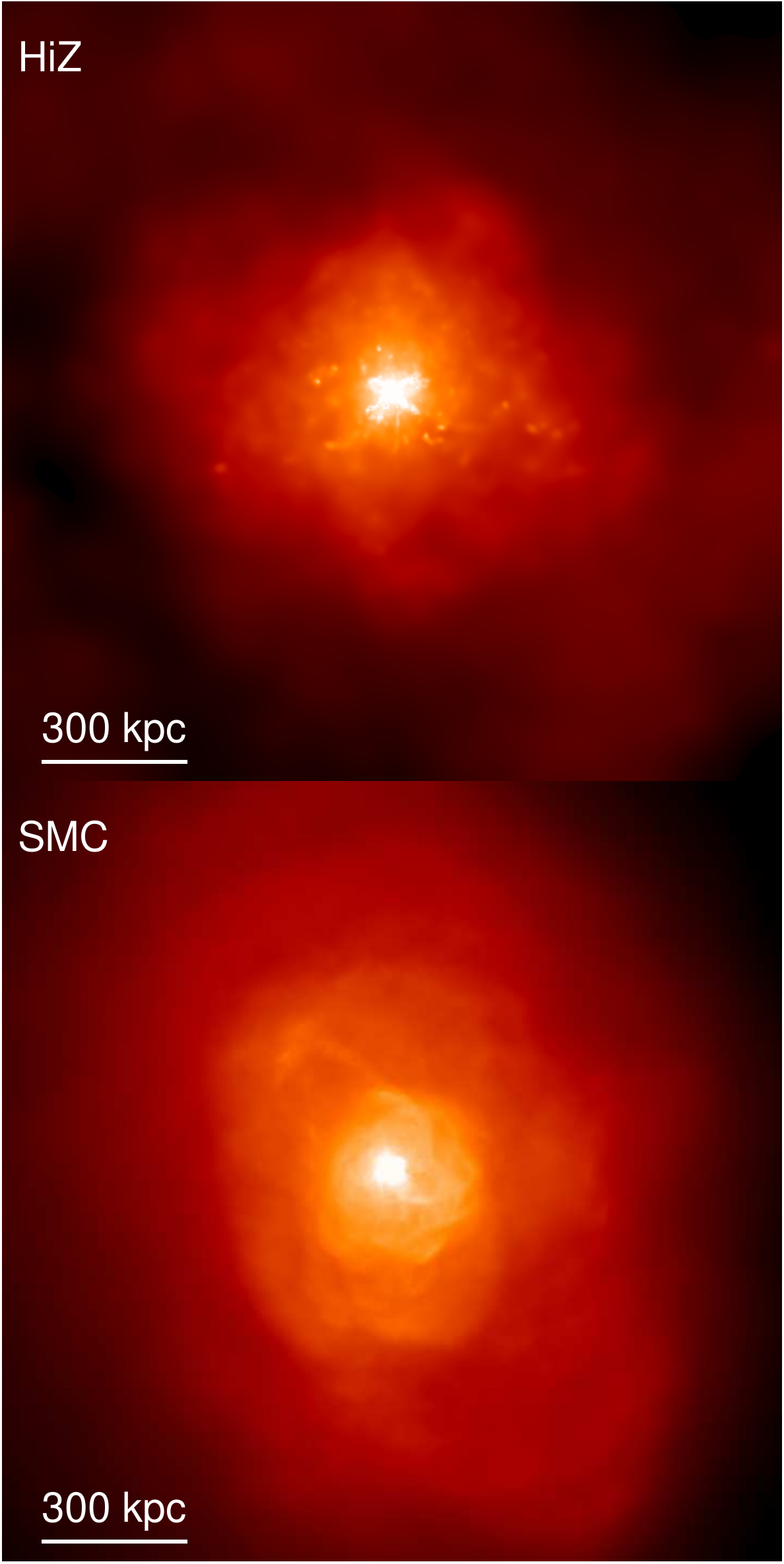}{0.6}
    \caption{Wind thermal+metal line emission morphology, as Fig.~\ref{fig:wind.images.xr}, but for the SMC {\bf f} merger re-run with the revised ``pressure-entropy'' formulation of SPH ({\em bottom}), as well as a HiZ {\bf e} merger run with the pressure-entropy SPH but lower resolution ($10x$ lower particle number). The large-scale wind behavior -- at least in these simulations which do not include a cosmological IGM -- is not strongly sensitive to the numerical method (modulo the small clumps in Fig.~\ref{fig:smc.mgr.morph}).
    \label{fig:wind.images.xr.vs.sph}}
\end{figure}

In this appendix, we discuss the robustness of the numerical methods used here -- in particular, we wish to study how the small-scale phase structure of the outflows can be affected by details of the methodology. 

Our default simulations in this paper use the standard ``density-entropy'' formulation of the SPH equations of motion in {\small GADGET} from \citet{springel:entropy}. This formulation manifestly conserves momentum, energy, angular momentum, and entropy (in the absence of sources/sinks), and has a number of additional advantages, but produces a resolution-scale ``surface tension''-like error term at contact discontinuities, which has the effect of suppressing the growth of some fluid mixing instabilities, and has been the subject of much discussion in the literature \citep[see][and references therein]{agertz:2007.sph.grid.mixing,read:2012.sph.w.dissipation.switches}. 

Since the multi-phase winds may well be subject to exactly these instabilities, we have re-run a subset of our simulations using the newer ``pressure-entropy'' SPH formulation described in \citet{hopkins:lagrangian.pressure.sph}, which is shown there to give dramatically improved results in situations with fluid mixing around contact discontinuities (e.g.\ the Kelvin-Helmholtz and Rayleigh-Taylor instabilities) while retaining excellent conservation properties, and includes a number of additional improvements to the treatment of artficial viscosity \citep[see][]{cullen:2010.inviscid.sph}, SPH smoothing kernel accuracy \citep{dehnen.aly:2012.sph.kernels}, and timestep communication relevant for treating extremely high Mach-number shocks \citep{saitoh.makino:2009.timestep.limiter,durier:2012.timestep.limiter}. For extensive numerical tests demonstrating accurate treatment of these instabilities, see \citet{hopkins:lagrangian.pressure.sph}. 

To test whether these subtleties may be strongly influencing our results, we first consider an isolated, star-forming disk (the progenitors in the SMC model mergers), which was analyzed in \citet{hopkins:lagrangian.pressure.sph}. In that paper, we ran otherwise exactly identical simulations (including the identical physical prescriptions to the merger simulations herein), but adopted either the density-entropy ``standard'' SPH or newer \citet{hopkins:lagrangian.pressure.sph} ``pressure-entropy'' form (in that case, keeping the kernel and all other properties fixed between the simulations). Fig.~15 in that paper compared the morphology of the isolated disks in those simulations. There we showed that the results were very similar; there were some small differences where the pressure-entropy formulation led to increased mixing along phase boundaries owing to the instabilities above, producing less-sharp divisions between molecular regions and hot bubbles. In that paper, we also compared the SFR and wind outflow rates as a function of time from the same isolated (SMC) disks. The time-averaged SFR differed only by $\sim20\%$. The total wind mass-loading was somewhat more strongly altered, and was lower in the pressure-entropy formulation by a factor of $\sim1.5-2$, because the increased mixing adds some cold gas to the hot medium which then substantially increases the hot gas cooling rate. So this suggests that the absolute wind mass-loading should be considered uncertain at the factor $\sim2$ level, in isolated disks.

Here, we extend this to compare a merger of these galaxies, our SMC {\bf f} model. Fig.~\ref{fig:smc.mgr.morph} shows the morphology of the winds at large radii in the merger (as Fig.~\ref{fig:wind.images}), for our standard ``density-entropy'' SPH formulation, and the newer ``pressure-entropy'' formulation. In the latter case, we now use the fully updated code from \citet{hopkins:lagrangian.pressure.sph}, with a more accurate artificial viscosity scheme, SPH kernel, and timestep limiter. In the newest version of the code, we have also implemented heating from the $z=0$ UV background as tabulated in \citet{faucher:ion.background.evol}, accounting for self-shielding, and we include this as well since it may well alter the phase distribution of the gas at large radii. The resulting differences at large radii are much more visually striking than those in the star-forming disk: the cold clumps or ``blobs'' at large radii disappear in the new simulation. We stress that the morphology of the smooth gas -- both the diffuse volume-filling hot gas and {\em also} the warm shells/filaments -- is nearly identical. And within the star-forming disk, GMCs still form in very similar fashion. It is largely these cold blobs at large radii that are altered. These are not self-gravitating, and are marginally resolved. They form by ``breakup'' of filamentary structures in the wind, in the density-entropy formulation. But this appears to be a numerical artifact of the density-entropy formulation, specifically the ``surface tension''-like term (which drives a beading effect along accelerating filaments that leads to breakup). The ionizing background suppresses their cooling to cold temperatures (though it has little effect on the morphology). 

These distinctions are more evident in the distribution of gas phases. In Fig.~\ref{fig:rho.dist.vs.sph}, we repeat Fig.~\ref{fig:rho.dist} and examine the density distribution of gas in the winds, ionized disk, and star-forming disk, as well as the specific phase breakdown within the star-forming disk. We compare both the isolated and merging SMC models with both density-entropy and pressure-entropy SPH. First, we emphasize that all of our qualitative conclusions appear robust. There are some quantitative changes, but most of these are in the tails of the density distributions, relevant at the sub-percent level. Within the star-forming disk, we see that the phase distributions are very similar in both implementations of SPH. Furthermore, including the ionizing background self-consistently has very little effect, except to truncate the cold gas density distribution at just about the density where our previous simple post-processing estimate led us to truncate the distributions. The predicted properties within the disk appear very robust. 

In Fig.~\ref{fig:wind.phases.vs.sph}, we extend this comparison to the velocity and density distributions of the wind material (as in Fig.~\ref{fig:wind.phases}). Here, we see larger differences, as expected from the previous Figures. In the pressure-entropy SPH, enhanced mixing of cold and hot phases decreases the {\em relative} importance of both in the wind, and enhances the relative importance of the warm-phase gas. The presence of an ionizing background also contributes to this. However, in each case we stress that warm-phase gas already dominated the outflow, so this conclusion is robust. The velocity distributions are altered, but only at a modest level -- there is still a broad velocity distribution in all phases (in fact, in the models here at the specific time analyzed, there may be somewhat more cold material at very large velocities, even though there is less overall in the outflow). 

We have specifically chosen to focus on the SMC case here, because its outflow, being predominantly ``hot phase'' gas (but featuring some cold blobs in the density-entropy runs) is most likely to be strongly affected by the details of the numerical method and fluid mixing. We have, however, also re-run lower-resolution versions of the HiZ {\bf e} and MW {\bf e} mergers with both the density-entropy and pressure-entropy formulations of SPH (and have experimented with a wide range of artificial viscosity and smoothing kernel implementations, as well as ionizing background strengths, in the isolated disk progenitors of each). In all these cases, we find the sense of the difference between SPH formulations is identical to that described above -- however, the quantitative magnitude of the difference is smaller in each case (see Fig.~\ref{fig:wind.images.xr.vs.sph}). The total wind mass-loading, for example, is only lower by $\sim20\%$ in the MW run, and actually appears to be slightly {\em higher} in the HiZ case. 

To summarize, the qualitative conclusions and results presented in the main text all appear robust to the details of the numerical method, {\em except} for the presence of cold ``blobs'' which form in the outflow. These features are likely driven by the same numerical artifacts that have already been identified as causing the breakup of inflows in cosmological simulations into similar ``clumpy'' morphologies \citep[see][]{keres:2011.arepo.gadget.disk.angmom,nelson:2013.arepo.coldflow.structure}. We stress, however, that these do not dominate the outflows by mass. Moreover, they do not appear until the outflow has essentially escaped the disk. At this point, as we have emphasized in the text, the detailed phase structure predicted should not be taken too seriously, since there is no IGM into which the winds expand. However, this should strongly caution against over-interpretation of the detailed phase distribution in cosmological simulations with outflows (many of which show similar features). 

Quantitatively, differences in the numerical method and ionizing background can lead to systematic changes in our predictions for basic wind properties (mass-loading factors, total mass in different phases, maximum velocities) at the factor of $\sim2$ level. This is significant. As shown in \paperthree, similar uncertainties can arise owing to the manner in which stellar feedback physics is implemented. And this is also comparable to genuine physical uncertainties, for example, the strength of feedback can vary at this level owing to plausible variation in the stellar initial mass function, or the presence of an equipartition magnetic field and/or cosmic rays \citep[see e.g.][and references therein]{uhlig:2012.cosmic.ray.streaming.winds,pakmor:2012.mag.field.disk.evol.weak.fx}. So we strongly emphasize that considerably more sophisticated simulations are needed (along with improved observational constraints) before any ``precision'' predictions with accuracy much better than a factor of $\sim2$ can be made.

\end{appendix}

\end{document}